\documentclass[%
article,
 amsmath,amssymb,
 twocolumn,
 superscriptaddress
]{revtex4}

\usepackage{graphicx}
\usepackage{dcolumn}
\usepackage{bm}

\begin{document}

\preprint{AIP/123-QED}

\title[Sample title]{A robust two-gene oscillator at the core of
  \emph{Ostreococcus tauri} circadian clock}

\author{Pierre-Emmanuel Morant}
\author{Quentin Thommen}
\author{Benjamin Pfeuty}
\author{Constant Vandermoere}
 \affiliation{Laboratoire de Physique des Lasers, Atomes, Mol\'ecules,
   Universit\'e Lille 1, CNRS, F-59655 Villeneuve d'Ascq, France}
 \affiliation{Institut de Recherche Interdisciplinaire,
   Universit\'e Lille 1, CNRS, F-59655 Villeneuve d'Ascq, France}
 \author{Florence Corellou}
 \author{Fran\c{c}ois-Yves Bouget}
 \affiliation{Laboratoire d'Oc\'eanographie Microbienne, Universit\'e
   Pierre et Marie Curie - Paris 6, CNRS, Observatoire Oc\'eanologique
   de Banyuls/Mer, 66650 Banyuls/Mer, France}
 \author{Marc Lefranc}
 \email{marc.lefranc@univ-lille1.fr}
 \affiliation{Laboratoire de Physique des Lasers, Atomes, Mol\'ecules,
   Universit\'e Lille 1, CNRS, F-59655 Villeneuve d'Ascq, France}
 \affiliation{Institut de Recherche Interdisciplinaire,
   Universit\'e Lille 1, CNRS, F-59655 Villeneuve d'Ascq, France}

\date{\today}

\begin{abstract}
  The microscopic green alga \emph{Ostreococcus tauri} is rapidly
  emerging as a promising model organism in the green lineage. In
  particular, recent results by Corellou \emph{et al.}
  [\emph{Plant Cell} \textbf{21}, 3436 (2009)]
  and Thommen \emph{et al.} [\emph{PLoS Comput. Biol.} \textbf{6},
    e1000990 (2010)] strongly suggest
  that its circadian clock is a simplified version of
  \emph{Arabidopsis thaliana} clock, and that it is architectured so
  as to be robust to natural daylight fluctuations. In this work, we
  analyze time series data from luminescent reporters for the two
  central clock genes \emph{TOC1} and \emph{CCA1} and correlate them
  with microarray data previously analyzed. Our mathematical analysis
  strongly supports
  both the existence of a simple two-gene oscillator at the core of
  \emph{Ostreococcus tauri} clock and the fact that its dynamics is
  not affected by light in normal entrainment conditions, a signature
  of its robustness.
\end{abstract}

\pacs{Valid PACS appear here}
\maketitle

\begin{quotation}
 \bf In order to anticipate periodic environmental
  changes induced by Earth rotation, many organisms have evolved a
  circadian clock, a genetic oscillator which generates biochemical rhythms
  with a period around 24 hours. Exact synchronization with the
  day/night cycle requires that one or more clock components sense
  daylight (for example, a protein degrades faster in the light).
  However, daylight intensity can highly fluctuate from day to day, or
  during the day, due to environmental factors such as sky cover. The
  question then arises as to how circadian clocks can keep time without
  being continously reset by the signal that should entrain them? A
  mathematical analysis of \emph{Ostreococcus tauri} clock, whose
  molecular basis has been identified
  recently~\cite{corellou09:_toc1_cca1}, has unveiled a simple and
  elegant mechanism which exploits the dynamical properties of the
  core clock oscillator to make it robust to daylight
  fluctuations~\cite{Thommen10}. When the clock is on time, coupling
  to light is activated precisely when the oscillator does not respond
  to external perturbations, making it blind to light and its
  fluctuations. If the clock has drifted and needs resetting, however,
  light affects the oscillator in a different part of its cycle, where
  it reacts so as to recover the entrainment phase. In this work, we
  provide strong evidence for the presence of a robust two-gene
  oscillator at the core of \emph{Ostreococcus} clock by showing that
  a minimal light-independent model can reproduce very accurately
  microarray data and luminescent reporter data recorded in different
  experiments.

\end{quotation}

\section{Introduction}

Biochemical oscillations are widespread biological phenomena involved
in many important cellular processes such as signaling, development,
motility or metabolism \cite{Goldbeter96book,Hasty10}.
Understanding how such a simple dynamical pattern has been implemented
repeatedly to support a great diversity of biological functions is
appealing for scientists seeking to uncover unifying
principles~\cite{Winfree01}. By identifying the molecular machinery
behind cellular rhythms, molecular biology has first laid the
groundwork for the development of strategies toward their
comprehensive understanding \cite{YoungMolGenBiolRhyth93}. In a second
stage, how a collective 
dynamics can emerge in networks of interacting molecular actors and how
it can be harnessed robustly has been under focus. Besides the design
of synthetic genetic circuits with specific oscillatory abilities
\cite{Elowitz00:_synth_net,Stricker08:_synth_osc}, a common strategy
to gain insight into this question has been to construct quantitative
dynamical models constrained by experiments
\cite{Goldbeter96book,Hartwell99,tiana07:_oscil,mengel10:_model_nf_b_wnt}. 
Indeed, the nonlinear dynamical behaviors that underlie oscillations can only
be fully captured through a mathematical description.

However, the ever-growing amount of experimental data obtained with
genetic transformations and real-time monitoring presents
extraordinary challenges to modelers. How to design relevant
mathematical models that not only reproduce the data but give insight
into the architecture of a biological system, and avoid pitfalls such
as overfitting which can ascribe biological meaning to experimental
artifacts? As we hope to illustrate here, it is important to combine
careful data analysis and minimal modeling, and to check the
consistence of the different data sets at all stages of model building.

Circadian clocks are systems of choice for quantitative studies of the
function and design of biochemical oscillators \cite{Rand04:_desig}.
They operate in many species to keep track of the most regular
environmental constraint, the alternation of daylight and darkness
caused by Earth rotation, so as to finely control the cellular
physiology accordingly~\cite{Pittendrigh60:_osc_circa}. This basic
function is vital in many organisms such as plants, which need to
timely coordinate their photosynthesis, and thus their growth and
division, to daily changes in light
intensity~\cite{Dodd05:_circa_osc,moulager07:_cell_divis_ostreoc,bouget10}.
However, a precise entrainment of circadian rhythms to the diurnal
cycle is potentially challenged
by many sources of variability including molecular
fluctuations~\cite{DidierGonze01222002,barkai00:_circadiannoise} and
temperature
variations~\cite{Pittendrigh54:_comp_temp,Rensing02:_comp_temp}, but  
also fluctuations of daylight
intensity from day to day or during the day, due to environmental
factors such as sky
cover~\cite{M.Comas10012008,Beersma08011999,Troein20091961,Thommen10}. 
To account for the remarkable ability of circadian oscillators to run
autonomously and to be precisely and robustly entrained, experimental
efforts aimed at unraveling their complex architecture
\cite{dunlap99:_molec,young01:_time,panda02:_circad} have motivated
studies trying to adjust mathematical models to experimental data
\cite{Forger12092003,forger:_mammalian,locke05:_extension,locke06:_exper,zeilinger06:_arabid_prr7_prr9,salazar09:_predic,Thommen10,francois05:_neurospora},
an approach that is increasingly necessary as models become more
complex, featuring generally several feedback loops.

In this paper, we use such a quantitative modeling approach to
investigate the dynamics of the circadian rhythms of the smallest
free-living eukaryote known to date, {\it Ostreococcus tauri}. This
microscopic green alga displays a very simple cellular organization
and a small and compact genome
\cite{Courties94:_ostreo_struc2,chretiennot95:_ostreo_struc,derelle06:_genome}.
Very recently, the molecular basis of its circadian clock has been
extensively characterized by Corellou \emph{et al.}, who carried
out an extensive work of genetic transformation, leading to
transcriptional and translational fusion lines allowing one to monitor
transcriptional activity and protein dynamics in living
cells~\cite{corellou09:_toc1_cca1,corellou10}. Their results point
to a core archictecture comprising two genes, similar to Arabidopsis
central clock genes \emph{TOC1} and \emph{CCA1}
\cite{DavidAlabadi08032001}. These two genes display rhythmic
expression both under light/dark alternation and in constant light
conditions. Thus, the unicellular green alga {\it Ostreococcus tauri}
has emerged as a promising organism model to study the circadian
rhythm in single photosynthetic eukaryotic cell combining experimental
and modeling approaches.

The goal of this modeling study is to check carefully whether
experimental data obtained through various channels support the
hypothesis that the circadian clock of {\it Ostreococcus Tauri}
contains a simple two-gene transcriptional loop serving as a core
oscillator~\cite{corellou09:_toc1_cca1,Thommen10}, and that the
dynamics of this oscillator is not affected by light in normal
entrainment conditions~\cite{Thommen10}. An important point of the
present work is the simultaneous adjustment of microarray data and
luminescent reporter data recorded in two different
experiments~\cite{moulager07:_cell_divis_ostreoc,corellou09:_toc1_cca1}
under different conditions. While this may not be sensible for a
general system whose parameters will typically change with
experimental conditions, it is expected here that the core clock
oscillator is robust to environmental changes. As advocated by Thommen
et al.~\cite{Thommen10}, it should thus deliver similar signals in
different experiments carried out with the same photoperiod.
Therefore, being able to reproduce with the same model signals
recorded with different techniques in different experiments provides
strong evidence that the core Ostreococcus clock oscillator is indeed
robust.

Before adjusting a minimal model to experimental data, a careful data
analysis of both microarray and luminescence time series has been
performed. It allowed us to detect and remove an experimental artifact
that would otherwise have prevented an optimal adjustment and thus
would have falsely called for a more complex model. The
combined use of microarray data and luminescence reporter data
provided a unique opportunity to calibrate the latter with respect to
the former, given that very little is known on luciferase dynamics in
Ostreococcus. This is all the more important as an extensive
collection of experimental data has been acquired with these reporters
and will form the basis of future quantitative studies of Ostreococcus
clock. 

We not only found that a simple two-gene transcriptional feedback loop
model reproduces perfectly the \emph{CCA1} and \emph{TOC1} transcript
and protein profiles obtained from data analysis, but that it does so
with no model parameter depending on light intensity whereas the data
have been recorded under light/dark alternation. This confirms our
previous observation based only on microarray data of limited time
resolution~\cite{Thommen10}. As we have proposed previously, this
counterintuitive finding indicates that the coupling to light does not
influence the oscillator in normal entrainment conditions, shielding
it from daylight fluctuations~\cite{Thommen10}. Besides confirming the
two-gene loop hypothesis and the oscillator robustness, this work
also identifies unambiguously the mechanistic origin of oscillation in the
transcriptional negative feedback loop involving \emph{CCA1} and
\emph{TOC1}, delayed by the saturated degradation of \emph{CCA1} mRNA and
TOC1 protein.

\section{Data Analysis}

In this section, we first present the experimental data and explain
how we determine mRNA and protein concentrations from them. We then
discuss the simplest model of the \emph{TOC1}--\emph{CCA1} negative
feedback loop, which consists of four differential equations
describing the time evolution of the \emph{TOC1} and \emph{CCA1} mRNA and protein
concentrations. To match the numerical time profiles generated by this
model, we need to reconstruct target profiles from the experimental
data. RNA profiles will be interpolated from microarray data whereas
protein profiles will be extracted from luminescence time series in
the form of a Fourier series describing the dynamics at long time
scales. Both types of profile will be characterized by their times of
passage at $20\%$ and $80\%$ of the oscillation amplitude, which
provide a uniform characterization of the two types of data considered here.
Model adjustment will be carried out by minimizing discrepancies
between the experimental and numerical passage times.

\subsection{Experimental data}

The microarray data used here come  from the study in
Ref.~\cite{moulager07:_cell_divis_ostreoc}. They were recorded at
three-hour time intervals under 12:12 light/dark alternation. The mRNA
time profiles that will serve as targets for model adjustment are the
same as in our previous study~\cite{Thommen10}, where we showed that
they could be accurately approached by solutions of a simple set of
differential equations, an indication of the high quality of these
data. The fact that microarray data reflect mRNA level without
ambiguity was also checked by quantitative RT-PCR.

The luminescent reporter data used here are those presented by
Corellou \emph{et al.} in~\cite{corellou09:_toc1_cca1} to provide
evidence of the central role of the \emph{TOC1} and \emph{CCA1} genes
in Ostreococcus clock. They had been recorded at one-hour intervals
under a 12:12 light/dark cycle from two types of transgenic cell
lines.

In transcriptional fusion lines pTOC1:luc and pCCA1:luc, the sequence
inserted into the genome consists of the promoter of one of the two
clock genes fused to the coding sequence of
firefly luciferase~\cite{corellou09:_toc1_cca1}, so that \emph{TOC1}
or \emph{CCA1} transcriptional activity drives luciferase expression.
Luciferase catalyzes the 
transformation of the substrate luciferin into oxyluciferin, in which
one photon is emitted \cite{Gates75}. The luminescence signal provides
information about the quantity of luciferase synthesized and thus on
the transcriptional activity of the promoters.

In translational fusion lines TOC1:luc and CCA1:luc, both the coding
sequences of one of the clock proteins (TOC1 or CCA1) and of
luciferase are fused to the promoter~\cite{corellou09:_toc1_cca1}, so
that a fusion protein combining the clock protein with luciferase is
synthesized by the additional gene with a trancriptional activity
similar to the original clock gene. In
this case, the luminescence signal also provides information about the
protein dynamics, and in particular on its degradation kinetics. It is
usually assumed that luciferase remains complexed and inactive after
the reaction and does not recover its activity before being degraded.
Thus two limiting cases can be considered depending on whether 
the luciferase reaction time (defined as the average time from
synthesis to photon emission) is much shorter or much longer that the
protein lifetime.

In the former case, luciferase reacts immediately after being
synthesized. Since only freshly made proteins then contribute to the
luminescence signal, the latter is obviously proportional to the
protein synthesis rate, hence to cytoplasmic fusion mRNA
concentration. If we assume for simplicity that the kinetic constants
(synthesis and degradation rates) of the fusion proteins TOC:luc and
CCA1:luc and of their mRNA are similar to that of their native
counterparts, then the concentrations of the former are proportional
to the concentrations of the latter. In this case, the luminescence
signal tracks mRNA concentration.

In the opposite case where the probability of photon emission by
luciferase is very low, and with the same assumption of identical
constants for fusion and native molecular actors, it is easily seen
that the luminescence signal is proportional to protein concentration.
In this limit, indeed, the photon emission probability is constant
throughout protein lifetime, thus luminescence intensity is
proportional to the number of TOC1:luc or CCA1:luc fusion proteins,
which is in turn proportional to the number of native clock proteins.
In this case the luminescence signal can be used as an indicator of
protein concentration. In the general case, it should be intermediate
between the RNA and protein time courses. 

A time lag of at least two hours between the maxima of RNA
concentration as indicated by microarray data and the maxima of
translational fusion lines can be observed, which indicates that
photon emission probability is low and that luciferase lifetime is
large compared to other protein lifetimes. This is consistent with
recent experiments in which translation was blocked in
\emph{Ostreococcus} cell cultures with emetine
dihydrochloride~\cite{corellou10} and time evolution of TOC:luc
luminescence was monitored. It was observed that the signal would
slowly decay over more than 12 hours, implying that the luciferase
reaction time is at least of this order. Thus, we will use in the
following translational fusion line signals as indicators of protein
concentrations, which allows us to keep our mathematical model as
simple as possible.

\subsection{Model of the \emph{TOC1}--\emph{CCA1} oscillator used for
  data adjustment}
The minimal model for the \emph{TOC1}--\emph{CCA1} transcriptional
feedback loop, where TOC1 activates \emph{CCA1} and CCA1
represses \emph{TOC1}, consists of the following four ordinary
differential equations:

  \begin{subequations}
    \label{eq:model}
  \begin{eqnarray}
    \dot{M_T} &=& \mu_T + \frac{\lambda_T}{1+(P_C/P_{C0})^{n_C}} -
    \delta_{M_T} \frac{K_{M_T} M_T}{K_{M_T} + M_T}\\
    \dot{P_T} &=& \beta_T M_T -
    \delta_{P_T} \frac{K_{P_T} P_T}{K_{P_T} + P_T}\\
    \dot{M_C} &=& \mu_C + \frac{\lambda_C (P_T/P_{T0})^{n_T}}{1+(P_T/P_{T0})^{n_T}} -
    \delta_{M_C} \frac{K_{M_C} M_C}{K_{M_C} + M_C}\\
    \dot{P_C} &=& \beta_C M_C -
    \delta_{P_C} \frac{K_{P_C} P_C}{K_{P_C} + P_C}
  \end{eqnarray}
\end{subequations}
Eqs.~(\ref{eq:model}) describe the time evolution of mRNA
concentrations $M_C$ and $M_T$ and protein concentrations $P_C$ and
$P_T$ for the \emph{CCA1} and \emph{TOC1} genes, respectively, as
they result from
mRNA synthesis regulated by the other protein, translation and
enzymatic degradation. \emph{TOC1} transcription rate varies between
$\mu_T$ at infinite CCA1 concentration and $\mu_T+\lambda_T$ at zero
CCA1 concentration according to the usual gene regulation function
with threshold $P_{C0}$ and cooperativity $n_C$. Similarly,
\emph{CCA1} transcription rate is $\mu_C$ (resp., $\mu_C+\lambda_C$)
at zero (resp., infinite) TOC1 concentration, with threshold $P_{T0}$
and cooperativity $n_T$. TOC1 and CCA1 translation rates are
$\beta_T$ and $\beta_C$, respectively. For each species $X$, the
Michaelis-Menten degradation term is written so that $\delta_X$ is the
low-concentration degradation rate and $K_X$ is the saturation
threshold.

A more detailed model would take into account compartmentalization,
with each actor having separate nuclear and cytoplasmic
concentrations, as well as the fact that the luminescence signal is
linked to a third gene artificially inserted in the genome, with the
kinetic constants of its mRNA and protein possibly different from
those of
its native counterpart. With this model, it is the profile predicted
for the total fusion protein concentration that would be adjusted to
the experimental time series rather than the native protein profile.
However, the important point is that such a biochemically detailed
model taking into account all native and inserted molecular actors
reduces to Eqs.~(\ref{eq:model}) in the limiting case where fusion
proteins and mRNA have the same kinetic constants as the native
molecules (e.g., TOC1:luc and TOC1 degradation rates are equal),
luciferase reaction time is large and nucleocytoplasmic transport is
fast. When the minimal model already adjusts the data with excellent
accuracy, as will be the case here, there is no point in using the
sophisticated model, which can only fit the data better. In fact, its
higher flexibility could lead to overfit the data and ascribe
incorrectly biological meaning to experimental artifacts.

As in~\cite{Thommen10}, the free-running period (FRP) of the uncoupled
oscillator is chosen equal to 24 hours, which was the mean value
observed in experiments~\cite{corellou09:_toc1_cca1}. In this case,
this oscillator is adjusted to experimental data without modulation
since our goal in this work is to show that the average oscillation
does not carry any signature of coupling to light. As a control, we
will also consider adjustment to oscillators of FRP 23.8 and 25 hours
under day conditions. In these two cases, a small modulation of some
parameters is required to achieve frequency locking. To keep the FRP
fixed during adjustment, we rescale kinetic parameters after each
optimization step so that the period of the free-running limit cycle
matches the desired value. This subsequently applies to the parameter
set which the adjustment converges to.

\subsection{Reconstruction of target profiles and determination of
  passage times}
\label{sec:reconstr-targ-prof}

\begin{figure}
  \centering
  \includegraphics[width=3.375in]{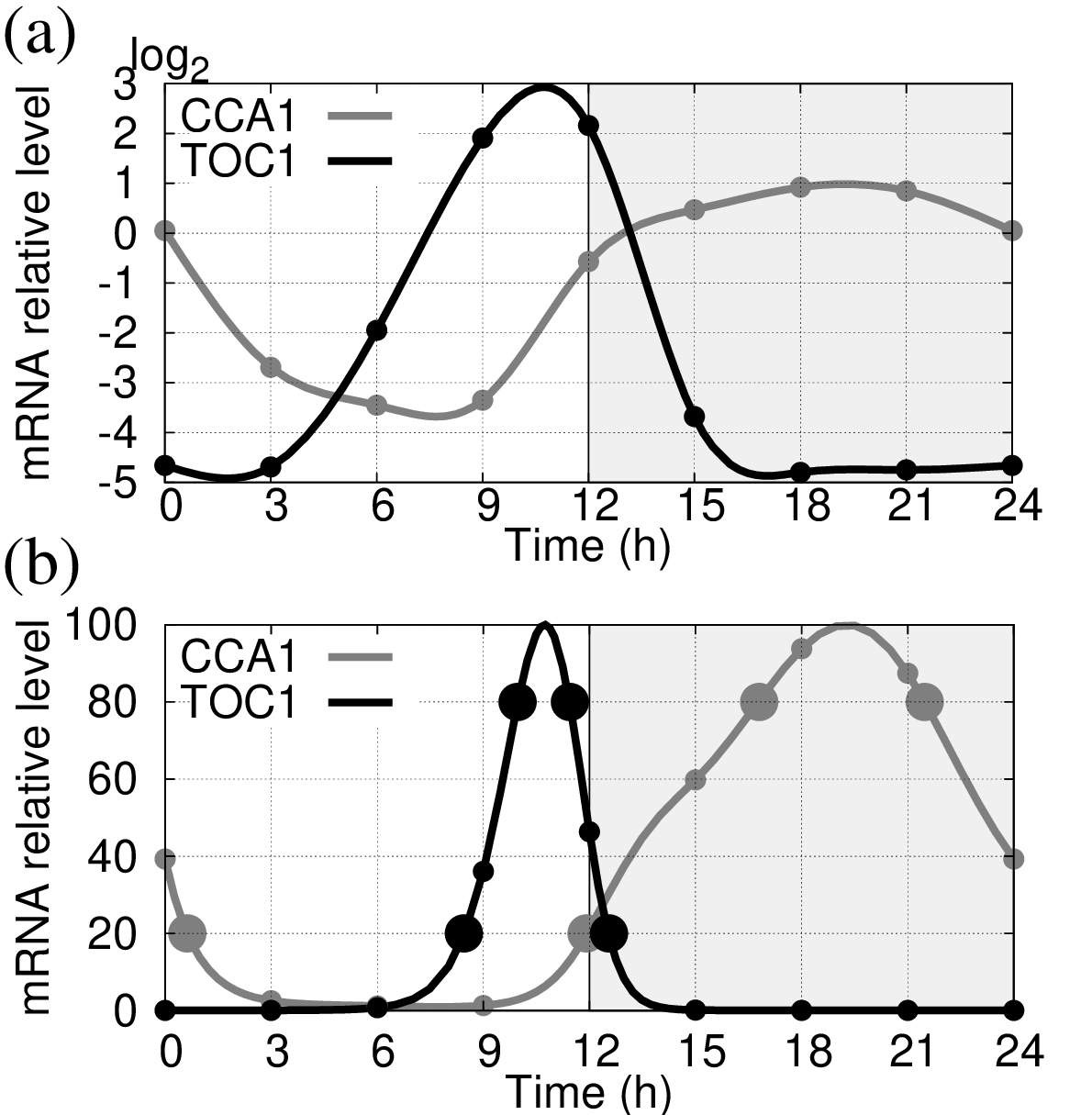}
  \caption{Construction of RNA target profiles as functions of
    Zeitgeber Time (ZT) describing phases within the dark/light cycle,
    with time ZT0 corresponds to dawn and time ZT12 to dusk.
    Interpolating curves going through data points (dots) are
    represented. (a) Interpolated curves from microarray data using
    cubic splines. (b) The exponentials of the interpolating curves
    were computed to obtain a smooth approximation of mRNA profiles.
    Large dots indicate passages through levels corresponding
    to 20\% and 80\% of maximum amplitude and serve as target points
    for adjustment.}
  \label{fig:rnatarget}
\end{figure}

Experimental RNA target profiles were estimated from the same
microarray data points as used in~\cite{Thommen10}. Due to the high
quality of data, noise reduction was not required. However, the
relatively low temporal resolution (one sample every three hours) made
it difficult to estimate accurately passage times and positions of
expression peaks. Thus the profiles were obtained by interpolating
between data points. Time series in the logarithm of RNA concentration
showed a very smooth behavior and were well approximated by simple
cubic splines (Fig.~\ref{fig:rnatarget}). These interpolating curves
were thus considered to provide an optimal approximation to the
variations of the logarithm of mRNA concentration, and mRNA linear
target curves were generated by exponentiation of the interpolating
curves. 

That the largest data point is less than half of the maximum of the
reconstructed \emph{TOC1} mRNA profile should not be surprising. In fact,
this profile is the most probable one given the data points if we
assume sufficient smoothness, or equivalently that the same dynamics
acts over the 24-hour cycle. Any other curve would imply fast,
transient, processes not linked to the \emph{TOC1}--\emph{CCA1} loop.
Since RNA concentration rises from almost zero at ZT6 (Zeitgeber Time
6, i.e., 6 hours after dawn) to almost the maximum level measured at
ZT9 and decays from the same level at ZT12 to almost zero at ZT15, it
is indeed obvious that the true maximum located between ZT9 and ZT12
must be much higher than the two largest data points. Incorrectly
assuming that the largest data point is the profile maximum would in
fact bias the analysis.

We checked that interpolating either the \emph{CCA1} mRNA concentrations or
their logarithms lead to two almost superimposing profile curves, showing
that both sequences of values provide the same information. However
the profile interpolated directly from \emph{TOC1} mRNA concentrations rather
than from their logarithms was highly nonphysical, with interpolated
concentrations becoming clearly negative on each side of the
expression peak. This is a direct consequence of the fact that only
two data points are well above the zero level. Implicit in the
interpolation procedure is the reconstruction of successive time
derivatives of a function from its values at different points. The
long sequence of almost zero data points makes the correspondence
between function values and time derivatives nearly singular, so that
the reconstruction is highly unstable. During the time interval near
zero, all time derivatives up to high order appear to be zero, then
jump to high values at the beginning and the end of the expression
peak. Since the interpolation error is proportional to the maximum
value of the fourth time derivate in the interval, the resulting
approximation is poor.

The series of logarithms of \emph{TOC1} mRNA
concentrations does not have this problem and allows optimal
reconstruction of the \emph{TOC1} mRNA profile. The fact that the logarithm
of a time series with long intervals near zero provides more
information about the dynamics than the original time series has also
been noted in the analysis of chaotic time
series~\cite{lefranc92:_correl_dimen,letellier10:_inter}. Furthermore,
it should be noted that many statistical analyses of microarray data
use the logarithms of mRNA concentrations because they are more evenly
distributed and provide more information.

Compared to a more usual method based on least-square-fitting the data
points (as used in~\cite{Thommen10}), this interpolation procedure can
only make adjustment more difficult, since it constrains tightly the
shape of the profile. However, it is fully consistent with the good
agreement found with a low-dimensional ODE set in our previous
study~\cite{Thommen10}. The passage times were then determined as the
times at which the interpolated profile would cross the appropriate
level (Fig.~\ref{fig:rnatarget}).

Protein profiles were reconstructed from translational fusion
luminescence signals used as indicators of protein concentrations, as
discussed above. The reconstruction was more involved than for mRNA
because these time series display significant amplitude variations
from peak to peak across the experiment, which may be for instance
linked to variations in the number of cells contributing to light
emission or other experimental factors.

\begin{figure}
  \centering
  \includegraphics[width=3.375in]{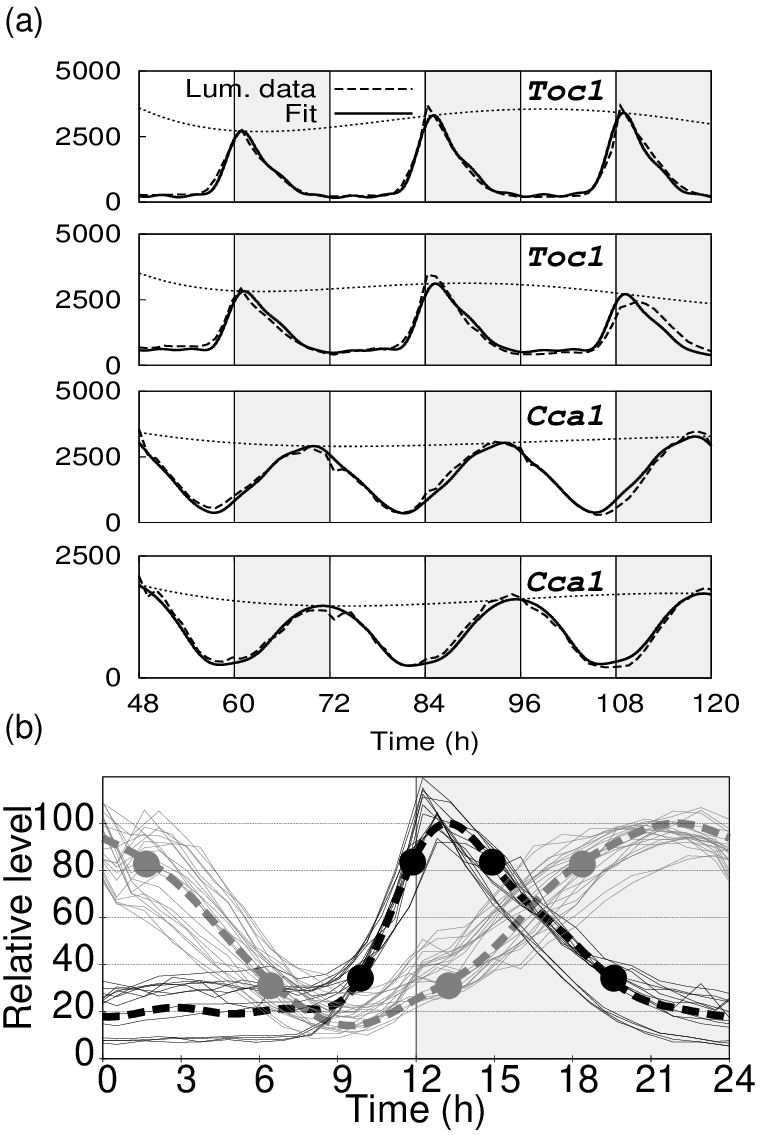}
  \caption{Reconstruction of protein target profiles as functions of
    Zeitgeber Time (ZT) describing phases within the dark/light cycle,
    with time ZT0 corresponds to dawn and time ZT12 to dusk. (a)
    Luminescence time series (dashed lines) for individual wells are
    fitted by the product of a Fourier series of period 24 hours with
    5 harmonics by a slowly varying polynomial function of degree 4
    (solid lines). The slowly varying envelope is also shown with a
    dotted line. From top to bottom, the first (last) two panels
    correspond to two TOC1:luc (CCA1:luc) translational fusion lines
    with different insertions in the genome. The
    different Fourier series are normalized to have the same maximum
    and are averaged. (b) The average Fourier series (dashed lines,
    black: TOC1, grey: CCA1) provides a smooth approximation to the
    cloud of individual line time profiles renormalized using the
    slowly varying polynomial function and wrapped around 24 hours
    (thin solid lines). Large dots indicate passages through levels
    corresponding to 20\% and 80\% of maximum amplitude and serve as
    target points for adjustment.}
  \label{fig:proteintarget}
\end{figure}

The first 48 recording hours, considered as transient, were discarded
and traces were then least-squares fitted by the product $p(t)F(t)$ of
a Fourier series $F(t)=\sum_{k=0}^{N}a_k \cos (2\pi k t/T+\phi_k)$,
with $T= 24$ hours, by a low-order polynomial $p(t)$ accounting for
slowly varying experimental conditions (Fig.~\ref{fig:proteintarget}).
The order of the polynomial $p(t)$ was generally chosen to be $3$ or
$4$, roughly equal to the number of periods in the fitted segment to
avoid over-fitting. The Fourier series $F(t)$ represents the average
periodic biochemical oscillation associated with circadian oscillations
while $p(t)$ models a slowly varying gain (due for example to a
variable number of cells contributing to light emission).

Because we are interested in long lasting mechanisms sustaining the
oscillation, the number $N$ of harmonics of the Fourier series was fixed
to 5 (an odd number, since the spectrum of the square wave forcing by
light only contains odd harmonics). We also tried slightly higher
numbers of harmonics, with little difference in the results. Fourier
fitting naturally smoothes out acute responses at day/night and
night/day transitions that can be seen in the raw signals and which
may reflect rapid mechanisms involved in clock resetting. This
separation of dynamical processes occurring on a 24-hour time scale on
the one hand, and confined to short time intervals on the other hand,
must be understood with reference to our previous finding that the
average dynamics of the \emph{TOC1}--\emph{CCA1} oscillator is
identical to that of a free-running oscillator, with coupling to light
and resetting occuring transiently in specific time
intervals~\cite{Thommen10}. Confirming this key observation with a
careful analysis of the luminescence time series is one of the main
goals of this paper. It will be reached if we find that the average
protein profile does not carry any signature of a coupling to light.

The Fourier series obtained from the fits of all bioluminescence
traces (possibly corresponding to different transgenic lines) recorded
in the experiment were then averaged, and the average was used as
surrogate for the protein temporal profile after suitable
normalization. The resulting profile is shown in
Fig.~\ref{fig:proteintarget}(b), along with individual luminescence
traces from different wells, renormalized using the slowly varying
polynomial function $p(t)$. The variability observed is partially due
to the fact that a few different transgenic lines, corresponding to
different insertions in the genome, were used. The times of passage at
$20\%$ and $80\%$ of the amplitude were then determined, as well as
the minimal level reached, for subsequent model adjustment.


We also tried fitting each luminescence trace to $p(t)F(t)+b$ where
$b$ represents a possible constant bias. Surprisingly, the fit was
consistently better when $b$ was equal to the minimum luminescence
level. In other words, the Fourier fitting procedure tended to remove
the floor level. Because this could indicate the existence of a bias
in the luminescence level, we examined more closely the raw data, and
found that indeed the zero luminescence level does not correspond to
the zero protein level.

\begin{figure}[!ht]
  \centering
  \includegraphics[width=8cm]{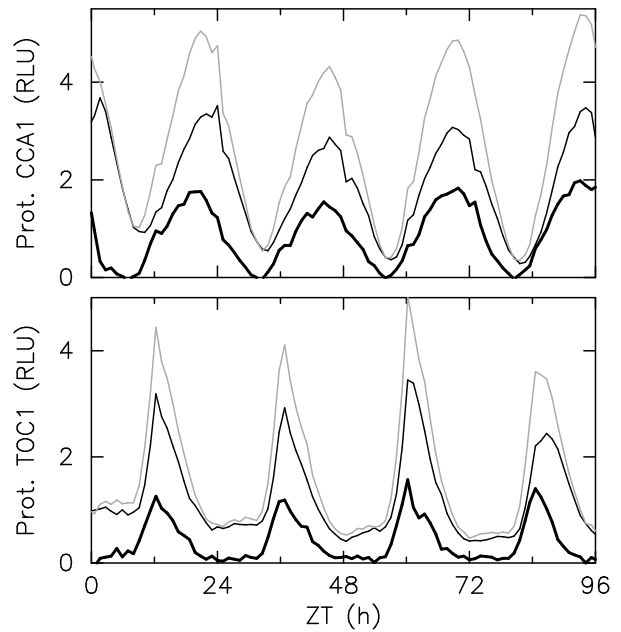}

  \caption{Translational fusion lines data recorded under 12:12 LD
    alternation. Time zero corresponds to dawn. On top panel (resp.,
    bottom panel) the time evolution of photon count (in relative
    luminescence units) of CCA1 (resp., TOC1) translational
    fusion lines are drawn for two biological duplicates monitored at
    the same time in the same conditions (black and grey thin solid
    lines). Their difference is plotted as a thick black solid
    line.}
   \label{fig:Biais}
\end{figure}

In Fig.~\ref{fig:Biais}, we show two individual traces monitoring
luminescence intensities over time in two wells of the luminometer
plate. These two wells contain genetically identical cell cultures
monitored simultaneously in the same experiment. Besides an excellent
overall reproducibility, the comparison of the two traces reveals the
existence of a significant bias. Assuming that identical clocks run in
different cells, and that there is a well-defined average number of
fusion proteins per cell, the luminescence signals should be
proportional to numbers of cells in each well. Therefore, the two
signals should be approximately proportional to each other.

Contrary to this, we observe that both for TOC1:luc and CCA1:luc, the maxima
of the two signals differ by a large factor, which remains roughly
constant over time while minima almost coincide. If the two traces
were proportional to each other, the minima should be in the same
ratio as the maxima. The only simple explanation for the systematic
coincidence of minima is that they correspond to a zero protein level,
and that there is the same bias on the two time series. This bias can
therefore be removed by substracting the two traces, as is shown in
Fig.~\ref{fig:Biais}, where the difference curves provide a better
estimation of the actual protein profiles than the original traces.
Importantly, the CCA1 protein level is predicted to touch zero very
shortly near ZT9, while the TOC1 protein level appears to stay near
zero for a large interval of time between ZT21 and ZT8
(Fig.~\ref{fig:Biais}). Note that this bias is not constant in time
but varies slowly so that the different minima correspond to different
luminescence levels.

While the existence of an experimental bias in the luminescence level
is strongly supported by the observations above, we have currently no
explanation for it. In the following, we will therefore adjust two
versions of the experimental profiles : (i) the Fourier series $F(t)$
determined directly from fitting $p(t)F(t)$ to the time series, as
described above, as if there was no bias, and (ii) the same Fourier
series with the floor level removed so that they touch zero, as with
the difference curves in Fig.~\ref{fig:Biais}. More precisely, the
target curves are then $F(t)-F_0$ where the $F(t)$ are obtained from
fitting $p(t)F(t)$ and $F_0 =\min_t F(t)$. This will allow us to
assess the influence of the bias on model fitting. Again, using a more
sophisticated model of the floor level bias could only lead to a
better adjustment, which will not be needed, and would only make sense with
a better understanding of the physical origin of the bias.

\subsection{Measuring goodness of fit by passage times}
\label{sec:meas-goodn-fit}

Adjusting experimental data to a mathematical model requires a
score function quantifying the discrepancy between experimental and
simulated profiles, to be minimized over parameter space. One
difficulty is that microarray and luminescent reporter data are very
different in nature and in particular have different uncertainties so that it
does make little sense to combine their adjustment errors. Therefore,
we used passage times as a unifying measure of fitting errors,
determining for each temporal profile the times of 
passage at $20\%$ and $80\%$ of the interval between the minimal and
the maximal values. We then compared passage times for the
experimental and numerical profiles, with the root mean square timing
error serving as a goodness of fit. Another advantage of passage times
is that they have a clear biological significance, with crossings of
the $20\%$ level bracketing the time interval with expression
significantly above background level and the crossings with the $80\%$
level bracketing the expression peak.

We denote by $\Delta^X_{20\uparrow}$ and $\Delta^X_{20\downarrow}$
(resp., $\Delta^X_{80\uparrow}$ and $\Delta^X_{80\downarrow}$) the time passage  
error in minutes for the concentration of species $X$ ($X=$ $M_C$,
$P_C$, $M_T$, $P_T$ ) at $20\%$ 
(resp., $80\%$) of the interval between minimal and maximal values
when increasing and decreasing.
For each species  $X$, one can therefore define a quadratic error function
equal to: 
\begin{equation}
Err(X)=\left(\Delta^X_{20\uparrow} \right)^2 +\left(\Delta^X_{80\uparrow}
  \right)^2+\left(\Delta^X_{80\downarrow}
  \right)^2+\left(\Delta^X_{20\downarrow} \right)^2
\end{equation}

The first, and most natural, score function is a RMS error combining
error for all species (mRNA and proteins):
\begin{equation}
S_{MP} =\sqrt{\frac{1}{16}\sum_{X\in {M_C,P_C,M_T,P_T}} Err(X)}
\end{equation}

In some cases, profiles with similar passage times differ by their
floor level.
We therefore introduce a floor level error
$\Phi_X=\frac{\min\left(\hat{X}\right)}{\max\left(\hat{X}\right)}-\frac{\min\left(X^*\right)}{\max\left(X^*\right)}$
where $\hat{X}$ is the numerical profile
for species $X$ and $X^*$ is the experimental target for $\hat{X}$. We choose a
ponderation such that a floor level error $\Phi_X=0.05$ is 
equivalent to a passage time error of 60 minutes. 
The score function of this second scheme is        
\begin{equation} 
S_{MPF}=\frac{1}{\sqrt{18}} \sqrt{ 16 \left( S_{MP}  \right)^2
+ \sum_{X\in {P_C,P_T}} \left(\Phi_{X}/\Phi_0\right)^2 }  \label{eq:mpf}
\end{equation}
where $\Phi_0=0.05/60=1/1200$ is the relative floor level error considered
equivalent to one minute.

Finally, microarray and luminescence data were not only adjusted
simultaneously but also 
separately in order to assess the coherence between the two types of
data. To this end, we use the following score functions:
\begin{eqnarray}
S_{M} &= & \sqrt{\frac{1}{8}\sum_{X\in {M_C,M_T}} Err(X)}\\
S_{P} &= & \sqrt{\frac{1}{10}\sum_{X\in {P_C,P_T}} \left(Err(X) + \left(\Phi_{X}/2\Phi_0 \right)^2 \right)}
\end{eqnarray}
where the $\Phi_X$ error term is scaled to $2\Phi_0$ so that it has
the same relative importance in the total RMS error as in
Eq.~(\ref{eq:mpf}). 

\section{Results: Adjustment of model to RNA and protein data}

\begin{figure*}[!ht]
  \centering
    \includegraphics[width=6in]{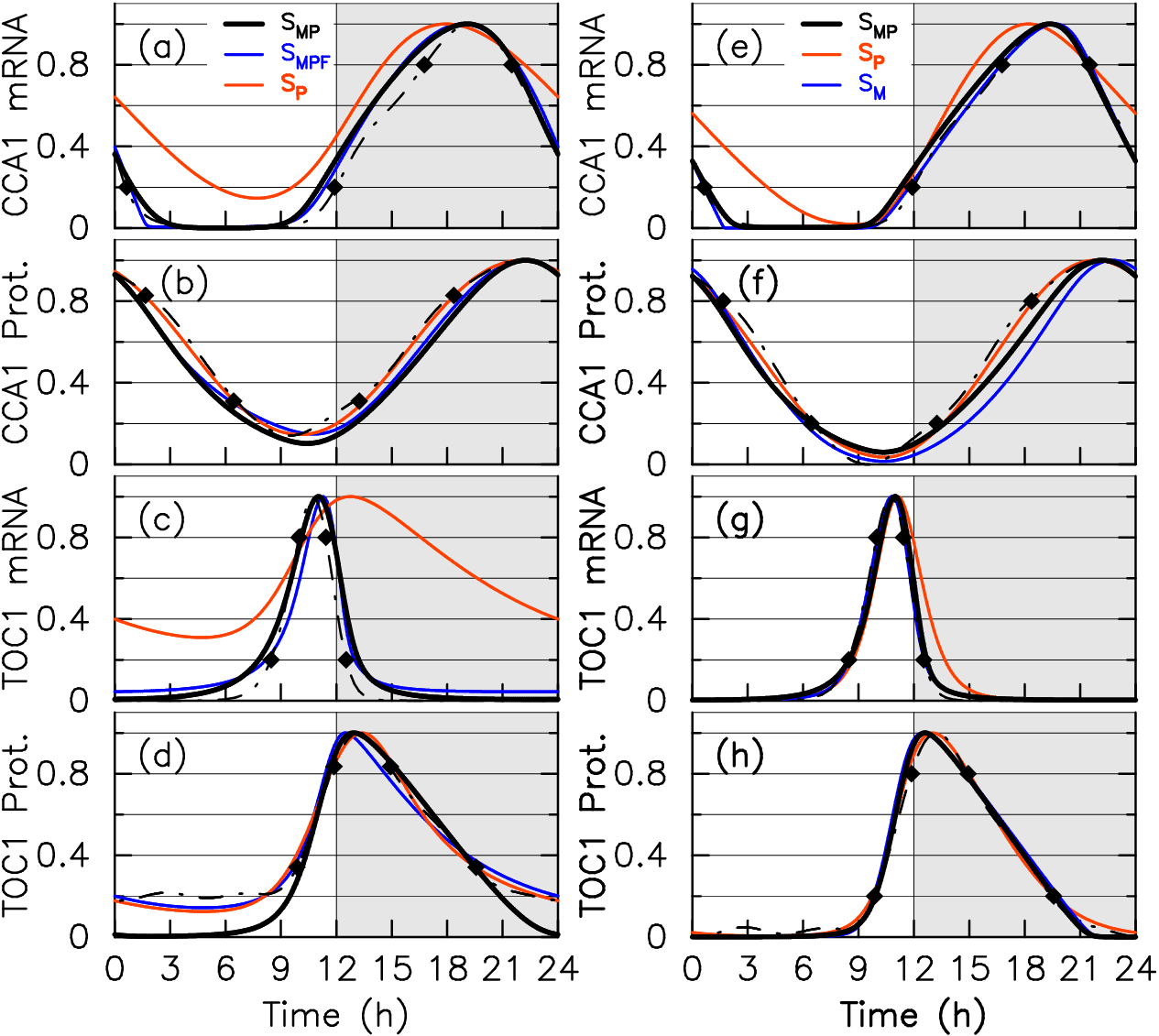}
    \caption{(Color figure) Adjustement of numerical solutions of the
      free-running model with FRP equal to 24~h (solid lines) to
      target curves (dash-dotted lines) using various score functions.
      Panels (a)-(d) in left column show adjustment of mRNA and
      protein profiles without bias correction, using score functions
      $S_{MP}$ (black solid line), $S_{MPF}$ (dark grey - blue thin
      solid line) and $S_P$ (light grey or red thin solid line). See
      text for the definition of score functions. Panels (e)-(h) in
      right column show adjustment of mRNA and protein profiles with
      protein floor level removed, using score functions $S_{MP}$
      (black solid line), $S_P$ (light grey - red thin solid line) and
      $S_M$ (dark grey or blue thin solid line). (a), (e): \emph{CCA1}
      mRNA; (b), (f) CCA1 protein; (c), (g) \emph{TOC1} mRNA; (d), (h)
      TOC1 protein. }
  \label{fig:ajust_24}
\end{figure*}

Figure~\ref{fig:ajust_24} shows the numerical solutions of
model~(\ref{eq:model}) best adjusting the experimental data according
to the score functions defined in Sec.~\ref{sec:meas-goodn-fit},
both without taking into account the floor level bias
(Figs.~\ref{fig:ajust_24}(a)-(d)) or with floor levels removed
(Figs.~\ref{fig:ajust_24}(e)-(h)). The corresponding parameter values
are given in Table~\ref{tab:PAR_24}.

\begin{table*}[!ht]\centering
  
  \caption{\textbf{Model parameter values}. Optimal parameter values for
    adjustement of model to data using various score functions, assuming a
    free-running period of $24$ hours. Parameters are rescaled so that the 
    maximum value of protein profiles is 100 nM, the maximum value of 
    \emph{CCA1} (resp., \emph{TOC1}) mRNA profile is 10 nM (resp., 70 nM). The
    \emph{TOC1} and \emph{CCA1} mRNA maximum values are chosen in the same
    proportion as in microarray data. The third row of the table indicates whether the floor levels of luminescence data are removed (R) or not (NR).
The last part of the table gives 
    the degradation
    rate $D_X$ at the mean value $\bar{X}=\left(
      \max(X)+\min(X)\right)/2$ ;
    $D_x=\delta_{X}K_{X}/\left(K_{X} + \bar{X} \right) $ for each species.}

 \begin{tabular}{ll|c|c|c|c|c|c}
                    &                              &           $S_{MP}$&          $S_{MPF}$&            $S_{P}$&            $S_{M}$&  $ S_{MP} $&    $S_{P}$\\ \hline
               score&                         (min)&     37.7&     32.8&     12.9&      4.2&     22.7&     11.3\\ \hline 
       Floor level &                              &     NR &     NR&     NR&      R/NR &     R&     R\\ \hline 
               $n_C$&                              &                   2&                   2&                   1&                   2&                   2&                   2\\ \hline
               $n_T$&                              &                   2&                   2&                   2&                   2&                   2&                   2\\ \hline
$\mu_C$             &(nM.$\textrm{h}^{-1}$)        &  1.92~$10^{-3}$&  1.21~$10^{-3}$&  5.18~$10^{-3}$&  2.97~$10^{-4}$&  1.53~$10^{-1}$&  1.46~$10^{-1}$\\ \hline 
$\lambda_C$         &(nM.$\textrm{h}^{-1}$)         &  3.64 &  6.61 &  6.56 &  3.32 &  3.11 &  3.31 \\ \hline 
$P_{T0}$           &(nM)                          &  31.6&  27.0&  91.6&  18.8&  18.7&  50.0\\ \hline 
$\beta_C$           &($\textrm{h}^{-1}$)           &  2.60 &  2.50 &  3.56 &  2.89 &  2.83 &  3.78 \\ \hline 
$\mu_T$             &(nM.$\textrm{h}^{-1}$)        &  0.0117&  104&  1.69 &  1.48 &  0.467&  0.0270\\ \hline 
$\lambda_T$         &(nM.$\textrm{h}^{-1}$)        &  560&  3130&  67.0&  272&  487&  233\\ \hline 
$P_{C0}$           &(nM)                          &  7.29 &  5.65 &  8.74 &  3.33 &  4.51 &  2.73 \\ \hline 
$\beta_T$           &($\textrm{h}^{-1}$)           &  0.667&  0.682&  6.86 &  0.811&  0.812&  0.759\\ \hline 
$1/\delta_{M_C}$    &(h)                           &  0.542&  0.0223&  0.831&  0.0161&  0.195&  0.652\\ \hline 
$1/\delta_{P_C}$    &(h)                           &  3.17 &  4.78 &  1.51 &  1.17 &  2.36 &  1.44 \\ \hline 
$1/\delta_{M_T}$    &(h)                           &  0.140&  0.0215&  4.97 &  0.151&  0.129&  0.736\\ \hline 
$1/\delta_{P_T}$    &(h)                           &  0.559&  4.20 &  0.0659&  0.118&  0.199&  1.65 \\ \hline 
$K_{M_C}$      &(nM)                          &  1.35 &  0.105&  2.24 &  0.0315&  0.407&  0.842\\ \hline 
$K_{P_C}$      &(nM)                          &  133&  746&  77.5&  23.6&  75.9&  72.2\\ \hline 
$K_{M_T}$      &(nM)                          &  37.7&  12.2&  128&  60.5&  28.3&  157\\ \hline 
$K_{P_T}$      &(nM)                          &  7.77 &  311 &  45.7&  1.52 &  2.76 &  47.6\\ \hline \hline
$D_{M_C}$           &($\textrm{h}^{-1}$)           &  0.220&  0.466&  0.220&  0.196&  0.201&  0.119\\ \hline 
$D_{P_C}$           &($\textrm{h}^{-1}$)           &  0.180&  0.185&  0.290&  0.164&  0.183&  0.292\\ \hline 
$D_{M_T}$           &($\textrm{h}^{-1}$)           &  2.49&  6.92&  0.130&  3.08&  2.23&  0.939\\ \hline 
$D_{P_T}$           &($\textrm{h}^{-1}$)           &  0.129&  0.180&  4.76&  0.127&  0.135&  0.196\\ \hline 
 \end{tabular}

  \label{tab:PAR_24}
\end{table*}

A first remark is that although a set of four passage times is a very
crude description of a temporal profile, it turns out that numerical
solutions and experimental profiles that have close passage times
generally follow each other closely all over the day. This is
consistent with the hypothesis that experimental data are well
described by a low-dimensional ODE set. Secondly, the quality of the
adjustment is globally very good given the simplicity of the model.
Clearly, Fig.~\ref{fig:ajust_24} shows that Eqs.~(\ref{eq:model}) can
adjust simultaneously the protein profiles reconstructed from
luminescence time series and the RNA profiles reconstructed from
microarray data (score functions $S_{MP}$ and $S_{MPF}$). Especially,
removing the protein profile floor levels so as to correct for the
detected experimental bias leads to impressive agreement between
numerical and target curves, which almost superimpose onto each other.

It is interesting to look more closely at the main discrepancies in
Figs~\ref{fig:ajust_24}(a)-(d). When only passage times are taken into
account (score function $S_{MP}$), the nonzero floor level of the TOC1
protein is not reproduced. This issue can be addressed by using a
score function that takes into account floor level error (score function
$S_{MPF}$). In this case, the floor level agreement is improved, at
the cost of slightly advancing the protein peak and of inducing a
small floor level for \emph{TOC1} mRNA. Although small, the latter is
definitely much higher than the value that can be accurately
determined from microarray data (which estimate the logarithm of mRNA
concentration). Last, the numerical solution obtained by adjusting the
protein profiles alone (score function $S_{P}$) predicts mRNA profiles
rather poorly (Fig.~\ref{fig:ajust_24}), which could suggest that the
two types of data are inconsistent.

To an uninformed mind, these minor discrepancies would most probably
suggest shortcomings of model~(\ref{eq:model}), not surprisingly given
its caricatural simplicity, or the fact that microarray and
luminescence data have been recorded in different experiments.
However, these discrepancies are clearly linked to errors in floor
level, which are naturally explained by the protein floor level bias
discussed in Sec.~\ref{sec:reconstr-targ-prof} and evidenced in
Fig.~\ref{fig:Biais}.

Let us now consider more closely adjustment of model~(\ref{eq:model})
to protein target profiles with floor levels removed, together with
RNA target profiles, shown in Figs.~\ref{fig:ajust_24}(e)-(h).
Goodness of fit is clearly improved compared to
Figs~\ref{fig:ajust_24}(a)-(d), where it was already good. The
simultaneous adjustment of the four target curves (score function
$S_{MP}$) is achieved with an accuracy rarely obtained for a genetic
circuit, with all numerical curves superimposing perfectly on the
experimental ones, except that the numerical CCA1 protein profile
rises slightly more slowly than the experimental one between dusk and
CCA1 expression peak. It is quite striking to note that the two curves
begin to separate precisely in the time interval where a transient
window of CCA1 stabilization was predicted to occur in
Ref.~\cite{Thommen10}. Remarkably, numerical profiles obtained by
adjusting RNA data alone (score function $S_M$) reproduce all
experimental data very well also. When protein data are adjusted
separately (score function $S_{P}$), RNA profiles are much better
reproduced than when the protein floor is not removed, with only the
\emph{CCA1} mRNA profile showing noticeable discrepancies in certain
parts of the 
day, but with the global shape of the profile preserved. Thus we can
conclude that removing floor level significantly improves adjustment
and restores coherence between microarray and luminescence data.
Given the simplicity of the model, it is very unlikely that this may
have occurred by chance. As we will discuss in the next section, it is
moreover a remarkable finding that experimental data recorded with
different techniques and in different
experiments~\cite{moulager07:_cell_divis_ostreoc,corellou09:_toc1_cca1}
can be adjusted simultaneously with such high accuracy by a minimal
model. This certainly reveals an important property of the clock
oscillator.

The parameter sets in the last three columns of
Table~\ref{tab:PAR_24}, which correspond to profiles with protein
floors removed, are generally more consistent between themselves than
those in the first three columns, where the bias has not been corrected,
and where important variations can be seen (compare for example
$\lambda_T$ across the first three columns). This reflects the fact
that when the floor level is not removed, it is more difficult to
predict mRNA profiles from adjusting protein profiles alone and vice
versa. To make meaningful comparisons, it should however be kept in
mind that the significant quantities are often not parameters
themselves but combinations of them. When degradation is saturated,
for example, the relevant quantity is the product $\delta_X K_X$, and
it may happen that $\delta_X$ and $K_X$ fluctuate more between columns
than their product. Therefore, we give at the bottom of
Table~\ref{tab:PAR_24} the effective degradation rates at a mean value
of the concentration $D_X$, which appear to be much
more consistent than the individual degradation parameters. Another
example is that although the values of $\mu_t$ in the fourth and fifth
columns are quite different, the minimum values taken by \emph{TOC1}
transcription rate in both cases are in fact comparable (1.78 vs 1.45
nM/h), and result from different combinations of $\lambda_T$,
$P_{C_0}$, and of the minimal value reached by the $P_C$ profile, which
is very close to the repression threshold $P_{C_0}$. This illustrates
the general fact that although collective fitting can provide
well-constrained predictions (here the time profiles), individual
parameters may be poorly constrained, a feature observed in many
systems biology models~\cite{Gutenkunst07}.

Some key ingredients of the dynamics can still be extracted
unambiguously from a careful examination of Table~\ref{tab:PAR_24}.
The first are the strongly saturated degradations of $M_C$ and $P_T$,
characterized by values of $K_{M_C}$ and $K_{P_T}$ so small that the
concentrations of the respective molecular actors are well above them
for almost all of the diurnal cycle. The signature of this behavior in
the time profiles is a straight-line decay after the expression peak.
A natural question is whether what appears as saturated degradation of
a given actor may in fact result from an interaction with another
actor not yet identified. In the case of TOC1, a post-transcriptional
regulation acting after dusk has indeed been evidenced
experimentally~\cite{corellou10}. Other key features are a maximum
transcription rate significantly higher for \emph{TOC1} than for
\emph{CCA1} and a small threshold of repression $P_{C_0}$, comparable
to the minimum value of the CCA1 protein level. These are related
since the latter implies that \emph{TOC1} is repressed most of the
time except in a very narrow time interval around the minimum of CCA1
protein level (this correlates with the narrow \emph{TOC1} mRNA peak).
This must 
be compensated for by a high \emph{TOC1} transcription rate.
Interestingly, the small value for $P_{C_0}$ can account for the
experimentally observed relative inefficiency of antisense strategies
against \emph{CCA1}~\cite{corellou09:_toc1_cca1}. If CCA1
level is reduced, even significantly, the small time interval where
\emph{TOC1} is not repressed will extend only slightly, with the
dynamical behavior at other times being mostly unchanged. Thus the
global perturbation will remain limited and arrhythmia will be
difficult to induce. In contrast to this, the activation threshold
$P_{T_0}$ above which TOC1 activates \emph{CCA1} is relatively low so
that activation occurs over a rather large time interval of more than
10 hours. Therefore, modifications of the activation threshold or TOC1
protein level will influence the dynamics for a long time and thus
easily disrupt oscillations. This also explains why FRP is much more
sensitive to overexpression of TOC1 induced by TOC1:luc insertion than
of CCA1 induced by CCA1:luc~\cite{corellou09:_toc1_cca1}.

\begin{figure}[ht]
  \centering
    \includegraphics[width=0.8\columnwidth]{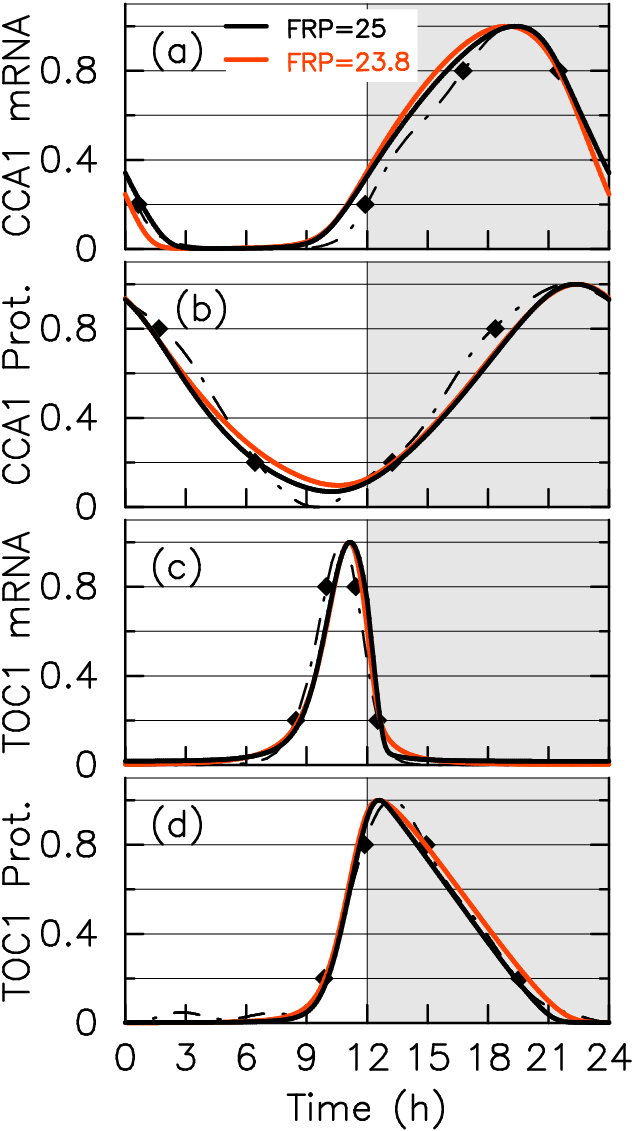}
  \caption{(Color figure) Adjustement of numerical solutions of the
    light-dependent model (solid lines) to target curves with floor levels removed (dash-dotted
    line) using the $S_{MP}$ score function.
    Adjustment of \emph{CCA1} mRNA (a), CCA1 protein (b),
    \emph{TOC1} mRNA (c), TOC1 protein (d) without the floor level of proteins.
    The FRP of the autonomous oscillator is either of 25~h (black solid line)
    and 23.8~h (light grey or red solid line). Synchronization is
    obtained for the 25~h (resp., 23.8~h) FRP model by assuming that
    the parameter $P_{C0}$ (resp., $\delta_{P_C}$) take a different value on the day and night.}
  \label{fig:ajust_FRP}
\end{figure}

Finally, we carried out data adjustment assuming free-running periods
of 23.8 and 25 hours, to verify that our results do not depend
critically on FRP (Fig.~\ref{fig:ajust_FRP}, parameters given in
Table~\ref{tab:PAR_FRP}). In this case, we allowed day/night
modulation of some parameters to ensure frequency locking with the
diurnal cycle. The agreement remains very good, although it degrades
noticeably compared to the case of a FRP of 24 hours. In particular, a
phase shift of the \emph{TOC1} mRNA profile is induced. The parameter
modulation 
remains very small (for a FRP of 25 hours, the repression threshold is
modulated but remains below the minimum of the CCA1 temporal profile,
so that the effect of the change is minimal). We furthermore observed
that when the FRP was a freely adjustable parameter, it would
systematically converge towards a value of 24 hours. All this confirms
that there is no day/night parameter modulation in the
\emph{TOC1}--\emph{CCA1} 
feedback loop and that coupling can only occur in specific time
windows as proposed in~\cite{Thommen10}.

\begin{table}[!ht]\centering
  
  \caption{\textbf{Model parameter values}. Optimal parameter values for adjustement of model to data using various score functions and assuming a free-running period of $25$ or $23.8$ hours. Parameters are rescaled as in Table \ref{tab:PAR_24}}    

 \begin{tabular}{ll|c|c}
                 FRP&                           (h)&                  25&                23.8\\ \hline
               score&                         (min)&     27.8&     34.3\\ \hline 
               $n_C$&                              &                   1&                   1\\ \hline
               $n_T$&                              &                   2&                   2\\ \hline
$\mu_C$             &(nM.$\textrm{h}^{-1}$)        &  6.06~$10^{-4}$&  4.73~$10^{-2}$\\ \hline 
$\lambda_C$         &(nM.$\textrm{h}^{-1}$)         &  3.82 &  4.99 \\ \hline 
$P_{TO}$           &(nM)                          &  15.9&  27.3\\ \hline 
$\beta_C$           &($\textrm{h}^{-1}$)           &  2.71 &  2.34 \\ \hline 
$\mu_T$             &(nM.$\textrm{h}^{-1}$)        &  22.7&  6.07~$10^{-2}$\\ \hline 
$\lambda_T$         &(nM.$\textrm{h}^{-1}$)        &  357&  1130\\ \hline 
$P_{C0}$           &(nM)                          &  5.94 &  6.15 \\ \hline 
$P_{C0}^\textrm{night}$&                       (nM)&  11.3&         \\ \hline 
$\beta_T$           &($\textrm{h}^{-1}$)           &  0.805&  0.809\\ \hline 
$1/\delta_{M_C}$    &(h)                           &  0.305&  0.296\\ \hline 
$1/\delta_{P_C}$    &(h)                           &  2.28 &  2.50 \\ \hline 
$1/\delta_{P_C}^\textrm{night}$&       (h)          &          &  2.60 \\ \hline 
$1/\delta_{M_T}$    &(h)                           &  4.28~$10^{-2}$&  4.51~$10^{-2}$\\ \hline 
$1/\delta_{P_T}$    &(h)                           &  0.224&  0.402\\ \hline 
$K_{M_C}$      &(nM)                          &  0.690&  0.879\\ \hline 
$K_{P_C}$      &(nM)                          &  68.5&  58.3\\ \hline 
$K_{M_T}$      &(nM)                          &  6.96 &  17.1\\ \hline 
$K_{P_T}$      &(nM)                          &  3.23 &  5.62 \\ \hline 
 \end{tabular}
  \label{tab:PAR_FRP}
  \end{table}

\section{Discussion and Conclusion}

In this work, we have carried out a careful and detailed analysis of
time series characterizing temporal variations of expression of the
two central genes of \emph{Ostreococcus} circadian clock, \emph{TOC1}
and \emph{CCA1}~\cite{corellou09:_toc1_cca1}. These time series had
been previously obtained from microarray data and luminescent reporter
data recorded in two different
experiments~\cite{moulager07:_cell_divis_ostreoc,corellou09:_toc1_cca1}.
From these time series, we have extracted periodic temporal profiles
for the RNA and protein concentrations of the two genes, assumed to
represent the mean circadian oscillations in individual cells so as to
optimize comparison with numerical profiles, which are inherently
periodic. In particular, approximating the periodic component of
luminescence time series by a Fourier series allowed us to separate
the dynamics at work throughout the day from fast transient processes
activated only near light/dark and dark/light transitions, and
probably involved in occasional resetting of the clock (compare raw
data and target profile in Fig.~\ref{fig:proteintarget}). The data
processing also allowed us to detect a bias in the luminescence time
series, which was confirmed by a direct comparison of individual
signals from two genetically identical cell cultures (they were not
proportional to each other). This bias manifests itself in a nonzero
luminescence floor level. Taking it into account allowed us to show
that both protein levels approach zero at some times of the day, which
was essential for the quality of the subsequent model adjustment. This
illustrates how important it is to exploit the redundancy of data by
verifying that data that should provide the same information are
indeed consistent.

Thanks to the careful data processing, we could evidence an
extraordinarily good agreement between a minimal model of a two-gene
transcriptional feedback loop and the reconstructed concentration
profiles. This model describes activation of \emph{CCA1} by
TOC1 and repression of \emph{TOC1} by CCA1. It describes
only regulated transcription of the two genes, translation and
degradation, and comprises only differential equations, from which the
time evolution of the two mRNA and the two proteins can be computed
and compared to the target profiles. Therefore a biochemically
detailed model taking into account compartmentalization or genomic
insertions would have only served to adjust details without biological
relevance, and could well have masked one of the main results which is
the good adjustment by a free-running oscillator model. It is also
quite remarkable that using only four points of the reconstructed
profiles sufficed to constrain the numerical profiles to follow their
targets throughout the day. Here also, adding more points would have
only forced the model to adjust to unrelevant details without gaining
information. This suggests that the complexity of the model and the
constraining data should be carefully matched.

The excellent agreement found between the model and the data supports
unambiguously the existence of a two-gene oscillator at the core of
\emph{Ostreococcus} circadian clock. This not only builds a solid
foundation on which future studies of this clock can rely but also
provides what we believe is one of the clearest examples of a natural
few-gene oscillator evidenced from experimental data. This is all the
more important as the role of this oscillator in the circadian clock
constrains it to be extremely robust to all kinds of fluctuations.
Understanding the dynamical ingredients besides transcriptional
regulation that underlie its robustness will certainly be of high
interest for the study of genetic oscillators in general. Two
remarkable features of the \emph{TOC1}--\emph{CCA1} oscillator have
indeed emerged from our analysis. First, a strongly saturated
degradation has been evidenced both for the \emph{CCA1} mRNA (already noted
in our previous work~\cite{Thommen10}) and the TOC1 protein (detected
in this work), which manifests itself by a straight-line decay at high
concentrations after the expression peak. This behavior may be the
signature of post-transcriptional and post-translation interactions,
and is thus compatible with the experimental observations of
Ref.~\cite{corellou10}. It also supports the putative role of
saturated degradation mechanisms as an efficient mechanism to
introduce effective delays along a negative feedback loop
\cite{Kurosawa02:_satur,tiana07:_oscil,Wong07:_prot_deg,degradator}
and the recent observation that this is a key ingredient to generate
robust oscillations~\cite{Stricker08:_synth_osc,Mather:_tsimringPRL}.
The second remarkable feature predicted by the model is that the
\emph{TOC1} gene is repressed by CCA1 during most of the day, except
during a short time interval located one or two hours before dusk, which is
consistent with the very narrow peak of \emph{TOC1} mRNA expression observed
in the experimental data. The small duration of \emph{TOC1} expression
is compensated by a high transcription rate. One may wonder whether
this design has an influence on the robustness to molecular
fluctuations. In any case, it explains why functional studies of the
oscillator showed that it was much more sensitive to pertubations in
the TOC1 level than in the CCA1 level~\cite{corellou09:_toc1_cca1}.

It has been recently been proposed that circadian clocks should not
only be robust to fluctuations in molecule
numbers~\cite{barkai00:_circadiannoise,DidierGonze01222002} and to
temperature
variations~\cite{Pittendrigh54:_comp_temp,Rensing02:_comp_temp} but
also to fluctuations in the daylight intensity pattern, which is
crucial to synchronize the circadian clock to the day/night
cycle~\cite{Troein20091961,Thommen10}. Our results demonstrate such
robustness for the \emph{TOC1}--\emph{CCA1} oscillator in two
different ways. First, the experimental data are reproduced accurately
by a free-running oscillator model. This had already been noted by
\citet{Thommen10}, but the confirmation of this behavior using
luminescence signals recorded every hour reinforces its plausibility
significantly. As shown in~\cite{Thommen10}, it can only be explained
by assuming that coupling to light is scheduled so as to be active
precisely when the oscillator does not respond to external
perturbations. In this scheme, the oscillator is not affected by light
in normal entrainment conditions, which naturally makes it blind to
light fluctuations, and thus behaves as if it was free-running. When
the oscillator drifts out of phase, light sensing occurs at a
different time of its cycle, where it responds so as to recover the
normal entrainment phase. The acute responses to light/dark and
dark/light transitions that occur transiently in raw signals
(Fig.~\ref{fig:proteintarget}) may be correspond to such resetting. 

The second strong evidence for the robustness of the
\emph{TOC1}--\emph{CCA1} oscillator comes from the fact that the
profiles adjusted precisely and simultaneously by our simple model are
reconstructed from time series recorded in different experiments by
two different techniques, with totally different setups and in
particular under very different lighting conditions. This strongly
suggests that \emph{Ostreococcus} clock is able to tick exactly in the
same way in different conditions, and in particular in different levels
of light, which is an obvious requisite 
for robustness to daylight fluctuations. If the clock can cope with
randomly varying daylight profiles without being perturbed, it can
certainly accomodate constantly low or constantly high daylight levels
and deliver similar biochemical signals in both cases. The fact that
the model adjusted from RNA profiles predicts protein profiles and
vice versa (although with less precision in the latter case) clearly
illustrates the consistency between the two experiments. Again, we
stress that this remarkable finding may have remained masked without
our careful data processing.

It is important to mention that both the presence of saturated degradation
kinetics and the absence of effective coupling to light when entrained are
strong predictions that can be tested experimentally.

While the impressive agreement between model and experiments obtained
here clearly shows that a \emph{TOC1}--\emph{CCA1} oscillator
underlies \emph{Ostreococcus} clock, it should not make us forget that
it is only a part of it. The free-running average behavior in
entrainment conditions, the acute responses at day/night and night/day
transitions can only be explained if the \emph{TOC1}--\emph{CCA1}
loop interacts with other molecular loops and feedback loops. Indeed
it was shown in~\cite{Thommen10} that robustness to daylight
fluctuations involves a precisely timed coupling activation. This
necessarily requires additional feedback loops designed so as to
generate the signal that will drive optimal coupling.
Furthermore, interactions with other molecular actors may well be
responsible for the saturated degradation of \emph{CCA1} mRNA and TOC1
protein. 

We thus have now to identify these additional feedback loops as well
as the light input pathways, which should help us to understand how
the clock can synchronize in spite of fluctuations to day/night cycles
of variable duration across the year. Adapting to different
photoperiods and to different light levels are indeed unrelated
evolutionary goals that must be simultaneously satisfied. In the end,
we hope to build an accurate and comprehensive model of a simple and
robust circadian clock. If the agreement between theory and experiment
remains comparable to what was achieved in the present study, this may
well provide us with new and deep insights into the design and
function of circadian clocks. More generally, we believe our results
promote Ostreococcus, whose low genomic
redundancy~\cite{derelle06:_genome} is probably crucial for allowing
accurate quantitative approaches, as a very promising model for
systems biology.

\section{ACKNOWLEDGMENTS}

This work has been supported by ANR grant 07BSYS004 to F-Y.B. and M.L., by
CNRS interdisciplinary programme ``Interface Physique, Biologie et Chimie:
soutien \`a la prise de risque'' to M.L., as well as by French Ministry
of Higher Education and Research, Nord-Pas de Calais 
Regional Council and FEDER through Contrat de Projets Etat-Region (CPER)
2007--2013.


\begin{thebibliography}{46}%
\makeatletter
\providecommand \@ifxundefined [1]{%
 \@ifx{#1\undefined}
}%
\providecommand \@ifnum [1]{%
 \ifnum #1\expandafter \@firstoftwo
 \else \expandafter \@secondoftwo
 \fi
}%
\providecommand \@ifx [1]{%
 \ifx #1\expandafter \@firstoftwo
 \else \expandafter \@secondoftwo
 \fi
}%
\providecommand \natexlab [1]{#1}%
\providecommand \enquote  [1]{``#1''}%
\providecommand \bibnamefont  [1]{#1}%
\providecommand \bibfnamefont [1]{#1}%
\providecommand \citenamefont [1]{#1}%
\providecommand \href@noop [0]{\@secondoftwo}%
\providecommand \href [0]{\begingroup \@sanitize@url \@href}%
\providecommand \@href[1]{\@@startlink{#1}\@@href}%
\providecommand \@@href[1]{\endgroup#1\@@endlink}%
\providecommand \@sanitize@url [0]{\catcode `\\12\catcode `\$12\catcode
  `\&12\catcode `\#12\catcode `\^12\catcode `\_12\catcode `\%12\relax}%
\providecommand \@@startlink[1]{}%
\providecommand \@@endlink[0]{}%
\providecommand \url  [0]{\begingroup\@sanitize@url \@url }%
\providecommand \@url [1]{\endgroup\@href {#1}{\urlprefix }}%
\providecommand \urlprefix  [0]{URL }%
\providecommand \Eprint [0]{\href }%
\providecommand \doibase [0]{http://dx.doi.org/}%
\providecommand \selectlanguage [0]{\@gobble}%
\providecommand \bibinfo  [0]{\@secondoftwo}%
\providecommand \bibfield  [0]{\@secondoftwo}%
\providecommand \translation [1]{[#1]}%
\providecommand \BibitemOpen [0]{}%
\providecommand \bibitemStop [0]{}%
\providecommand \bibitemNoStop [0]{.\EOS\space}%
\providecommand \EOS [0]{\spacefactor3000\relax}%
\providecommand \BibitemShut  [1]{\csname bibitem#1\endcsname}%
\let\auto@bib@innerbib\@empty
\bibitem [{\citenamefont {Corellou}\ \emph {et~al.}(2009)\citenamefont
  {Corellou}, \citenamefont {Schwartz}, \citenamefont {Motta}, \citenamefont
  {Djouani-Tahri}, \citenamefont {Sanchez},\ and\ \citenamefont
  {Bouget}}]{corellou09:_toc1_cca1}%
  \BibitemOpen
  \bibfield  {author} {\bibinfo {author} {\bibfnamefont {F.}~\bibnamefont
  {Corellou}}, \bibinfo {author} {\bibfnamefont {C.}~\bibnamefont {Schwartz}},
  \bibinfo {author} {\bibfnamefont {J.-P.}\ \bibnamefont {Motta}}, \bibinfo
  {author} {\bibfnamefont {E.~B.}\ \bibnamefont {Djouani-Tahri}}, \bibinfo
  {author} {\bibfnamefont {F.}~\bibnamefont {Sanchez}}, \ and\ \bibinfo
  {author} {\bibfnamefont {F.-Y.}\ \bibnamefont {Bouget}},\ }\href@noop {}
  {\bibfield  {journal} {\bibinfo  {journal} {Plant Cell}\ }\textbf {\bibinfo
  {volume} {21}},\ \bibinfo {pages} {3436} (\bibinfo {year}
  {2009})}\BibitemShut {NoStop}%
\bibitem [{\citenamefont {Thommen}\ \emph {et~al.}(2010)\citenamefont
  {Thommen}, \citenamefont {Pfeuty}, \citenamefont {Morant}, \citenamefont
  {Correlou}, \citenamefont {Bouget},\ and\ \citenamefont
  {Lefranc}}]{Thommen10}%
  \BibitemOpen
  \bibfield  {author} {\bibinfo {author} {\bibfnamefont {Q.}~\bibnamefont
  {Thommen}}, \bibinfo {author} {\bibfnamefont {B.}~\bibnamefont {Pfeuty}},
  \bibinfo {author} {\bibfnamefont {P.}~\bibnamefont {Morant}}, \bibinfo
  {author} {\bibfnamefont {F.}~\bibnamefont {Correlou}}, \bibinfo {author}
  {\bibfnamefont {F.}~\bibnamefont {Bouget}}, \ and\ \bibinfo {author}
  {\bibfnamefont {M.}~\bibnamefont {Lefranc}},\ }\href@noop {} {\bibfield
  {journal} {\bibinfo  {journal} {Plos Comput. Biol.}\ }\textbf {\bibinfo
  {volume} {6}},\ \bibinfo {pages} {e1000990. doi:10.1371/journal.pcbi.1000990}
  (\bibinfo {year} {2010})}\BibitemShut {NoStop}%
\bibitem [{\citenamefont {Goldbeter}(1996)}]{Goldbeter96book}%
  \BibitemOpen
  \bibfield  {author} {\bibinfo {author} {\bibfnamefont {A.}~\bibnamefont
  {Goldbeter}},\ }\href@noop {} {\emph {\bibinfo {title} {Biochemical
  Oscillations and Cellular Rhythms}}}\ (\bibinfo  {publisher} {Cambridge
  University Press},\ \bibinfo {address} {Cambridge},\ \bibinfo {year}
  {1996})\BibitemShut {NoStop}%
\bibitem [{\citenamefont {Hasty}, \citenamefont {Hoffmann},\ and\ \citenamefont
  {Golden}(2010)}]{Hasty10}%
  \BibitemOpen
  \bibfield  {author} {\bibinfo {author} {\bibfnamefont {J.}~\bibnamefont
  {Hasty}}, \bibinfo {author} {\bibfnamefont {A.}~\bibnamefont {Hoffmann}}, \
  and\ \bibinfo {author} {\bibfnamefont {S.}~\bibnamefont {Golden}},\
  }\href@noop {} {\bibfield  {journal} {\bibinfo  {journal} {Curr. Opin. Genet.
  Dev.}\ }\textbf {\bibinfo {volume} {20}},\ \bibinfo {pages} {571} (\bibinfo
  {year} {2010})}\BibitemShut {NoStop}%
\bibitem [{\citenamefont {Winfree}(2001)}]{Winfree01}%
  \BibitemOpen
  \bibfield  {author} {\bibinfo {author} {\bibfnamefont {A.~T.}\ \bibnamefont
  {Winfree}},\ }\href@noop {} {\emph {\bibinfo {title} {The Geometry of
  Biological Time}}},\ \bibinfo {edition} {3rd}\ ed.\ (\bibinfo  {publisher}
  {Springer Verlag, Berlin},\ \bibinfo {year} {2001})\BibitemShut {NoStop}%
\bibitem [{\citenamefont {Young}(1993)}]{YoungMolGenBiolRhyth93}%
  \BibitemOpen
  \bibinfo {editor} {\bibfnamefont {M.~W.}\ \bibnamefont {Young}},\ ed.,\
  \href@noop {} {\emph {\bibinfo {title} {Molecular Genetics of Biological
  Rhythms}}}\ (\bibinfo  {publisher} {Dekker, New York},\ \bibinfo {year}
  {1993})\BibitemShut {NoStop}%
\bibitem [{\citenamefont {Elowitz}\ and\ \citenamefont
  {Leibler}(2000)}]{Elowitz00:_synth_net}%
  \BibitemOpen
  \bibfield  {author} {\bibinfo {author} {\bibfnamefont {M.}~\bibnamefont
  {Elowitz}}\ and\ \bibinfo {author} {\bibfnamefont {S.}~\bibnamefont
  {Leibler}},\ }\href@noop {} {\bibfield  {journal} {\bibinfo  {journal}
  {Nature}\ }\textbf {\bibinfo {volume} {403}},\ \bibinfo {pages} {335}
  (\bibinfo {year} {2000})}\BibitemShut {NoStop}%
\bibitem [{\citenamefont {Stricker}\ \emph {et~al.}(2008)\citenamefont
  {Stricker}, \citenamefont {Cookson}, \citenamefont {Bennett}, \citenamefont
  {Mather}, \citenamefont {Tsimring},\ and\ \citenamefont
  {Hasty}}]{Stricker08:_synth_osc}%
  \BibitemOpen
  \bibfield  {author} {\bibinfo {author} {\bibfnamefont {J.}~\bibnamefont
  {Stricker}}, \bibinfo {author} {\bibfnamefont {S.}~\bibnamefont {Cookson}},
  \bibinfo {author} {\bibfnamefont {M.~R.}\ \bibnamefont {Bennett}}, \bibinfo
  {author} {\bibfnamefont {W.~H.}\ \bibnamefont {Mather}}, \bibinfo {author}
  {\bibfnamefont {L.~S.}\ \bibnamefont {Tsimring}}, \ and\ \bibinfo {author}
  {\bibfnamefont {J.}~\bibnamefont {Hasty}},\ }\href@noop {} {\bibfield
  {journal} {\bibinfo  {journal} {Nature}\ }\textbf {\bibinfo {volume} {456}},\
  \bibinfo {pages} {516} (\bibinfo {year} {2008})}\BibitemShut {NoStop}%
\bibitem [{\citenamefont {Hartwell}\ \emph {et~al.}(1999)\citenamefont
  {Hartwell}, \citenamefont {Hopfield}, \citenamefont {Leibler},\ and\
  \citenamefont {Murray}}]{Hartwell99}%
  \BibitemOpen
  \bibfield  {author} {\bibinfo {author} {\bibfnamefont {L.}~\bibnamefont
  {Hartwell}}, \bibinfo {author} {\bibfnamefont {J.}~\bibnamefont {Hopfield}},
  \bibinfo {author} {\bibfnamefont {S.}~\bibnamefont {Leibler}}, \ and\
  \bibinfo {author} {\bibfnamefont {A.}~\bibnamefont {Murray}},\ }\href@noop {}
  {\bibfield  {journal} {\bibinfo  {journal} {Nature}\ }\textbf {\bibinfo
  {volume} {402}},\ \bibinfo {pages} {47} (\bibinfo {year} {1999})}\BibitemShut
  {NoStop}%
\bibitem [{\citenamefont {Tiana}\ \emph {et~al.}(2007)\citenamefont {Tiana},
  \citenamefont {Krishna}, \citenamefont {Pigolotti}, \citenamefont {Jensen},\
  and\ \citenamefont {Sneppen}}]{tiana07:_oscil}%
  \BibitemOpen
  \bibfield  {author} {\bibinfo {author} {\bibfnamefont {G.}~\bibnamefont
  {Tiana}}, \bibinfo {author} {\bibfnamefont {S.}~\bibnamefont {Krishna}},
  \bibinfo {author} {\bibfnamefont {S.}~\bibnamefont {Pigolotti}}, \bibinfo
  {author} {\bibfnamefont {M.~H.}\ \bibnamefont {Jensen}}, \ and\ \bibinfo
  {author} {\bibfnamefont {K.}~\bibnamefont {Sneppen}},\ }\href@noop {}
  {\bibfield  {journal} {\bibinfo  {journal} {Phys. Biol.}\ }\textbf {\bibinfo
  {volume} {4}},\ \bibinfo {pages} {R1} (\bibinfo {year} {2007})}\BibitemShut
  {NoStop}%
\bibitem [{\citenamefont {Mengel}\ \emph {et~al.}(2010)\citenamefont {Mengel},
  \citenamefont {Hunziker}, \citenamefont {Pedersen}, \citenamefont {Trusina},
  \citenamefont {Jensen},\ and\ \citenamefont
  {Krishna}}]{mengel10:_model_nf_b_wnt}%
  \BibitemOpen
  \bibfield  {author} {\bibinfo {author} {\bibfnamefont {B.}~\bibnamefont
  {Mengel}}, \bibinfo {author} {\bibfnamefont {A.}~\bibnamefont {Hunziker}},
  \bibinfo {author} {\bibfnamefont {L.}~\bibnamefont {Pedersen}}, \bibinfo
  {author} {\bibfnamefont {A.}~\bibnamefont {Trusina}}, \bibinfo {author}
  {\bibfnamefont {M.~H.}\ \bibnamefont {Jensen}}, \ and\ \bibinfo {author}
  {\bibfnamefont {S.}~\bibnamefont {Krishna}},\ }\href@noop {} {\bibfield
  {journal} {\bibinfo  {journal} {Curr. Opon. Genet. Dev.}\ }\textbf {\bibinfo
  {volume} {20}},\ \bibinfo {pages} {656} (\bibinfo {year} {2010})}\BibitemShut
  {NoStop}%
\bibitem [{\citenamefont {Rand}\ \emph {et~al.}(2004)\citenamefont {Rand},
  \citenamefont {Shulgin}, \citenamefont {Salazar},\ and\ \citenamefont
  {Millar}}]{Rand04:_desig}%
  \BibitemOpen
  \bibfield  {author} {\bibinfo {author} {\bibfnamefont {D.~A.}\ \bibnamefont
  {Rand}}, \bibinfo {author} {\bibfnamefont {B.~V.}\ \bibnamefont {Shulgin}},
  \bibinfo {author} {\bibfnamefont {D.}~\bibnamefont {Salazar}}, \ and\
  \bibinfo {author} {\bibfnamefont {A.~J.}\ \bibnamefont {Millar}},\
  }\href@noop {} {\bibfield  {journal} {\bibinfo  {journal} {J. R. Soc.
  Interface}\ }\textbf {\bibinfo {volume} {1}},\ \bibinfo {pages} {119}
  (\bibinfo {year} {2004})}\BibitemShut {NoStop}%
\bibitem [{\citenamefont {Pittendrigh}(1960)}]{Pittendrigh60:_osc_circa}%
  \BibitemOpen
  \bibfield  {author} {\bibinfo {author} {\bibfnamefont {C.~S.}\ \bibnamefont
  {Pittendrigh}},\ }\href@noop {} {\bibfield  {journal} {\bibinfo  {journal}
  {Cold Spring Harb. Symp. Quant. Biol.}\ }\textbf {\bibinfo {volume} {25}},\
  \bibinfo {pages} {159} (\bibinfo {year} {1960})}\BibitemShut {NoStop}%
\bibitem [{\citenamefont {Dodd}\ \emph {et~al.}(2005)\citenamefont {Dodd},
  \citenamefont {Salathia}, \citenamefont {Hall}, \citenamefont {Kevei},
  \citenamefont {Toth}, \citenamefont {Nagy}, \citenamefont {Hibberd},
  \citenamefont {Millar},\ and\ \citenamefont {Webb}}]{Dodd05:_circa_osc}%
  \BibitemOpen
  \bibfield  {author} {\bibinfo {author} {\bibfnamefont {A.~N.}\ \bibnamefont
  {Dodd}}, \bibinfo {author} {\bibfnamefont {N.}~\bibnamefont {Salathia}},
  \bibinfo {author} {\bibfnamefont {A.}~\bibnamefont {Hall}}, \bibinfo {author}
  {\bibfnamefont {E.}~\bibnamefont {Kevei}}, \bibinfo {author} {\bibfnamefont
  {R.}~\bibnamefont {Toth}}, \bibinfo {author} {\bibfnamefont {F.}~\bibnamefont
  {Nagy}}, \bibinfo {author} {\bibfnamefont {J.}~\bibnamefont {Hibberd}},
  \bibinfo {author} {\bibfnamefont {A.~J.}\ \bibnamefont {Millar}}, \ and\
  \bibinfo {author} {\bibfnamefont {A.~A.}\ \bibnamefont {Webb}},\ }\href@noop
  {} {\bibfield  {journal} {\bibinfo  {journal} {Science}\ }\textbf {\bibinfo
  {volume} {309}},\ \bibinfo {pages} {630} (\bibinfo {year}
  {2005})}\BibitemShut {NoStop}%
\bibitem [{\citenamefont {Moulager}\ \emph {et~al.}(2007)\citenamefont
  {Moulager}, \citenamefont {Monnier}, \citenamefont {Jesson}, \citenamefont
  {Bouvet}, \citenamefont {Mosser}, \citenamefont {Schwartz}, \citenamefont
  {Garnier}, \citenamefont {Corellou},\ and\ \citenamefont
  {Bouget}}]{moulager07:_cell_divis_ostreoc}%
  \BibitemOpen
  \bibfield  {author} {\bibinfo {author} {\bibfnamefont {M.}~\bibnamefont
  {Moulager}}, \bibinfo {author} {\bibfnamefont {A.}~\bibnamefont {Monnier}},
  \bibinfo {author} {\bibfnamefont {B.}~\bibnamefont {Jesson}}, \bibinfo
  {author} {\bibfnamefont {R.}~\bibnamefont {Bouvet}}, \bibinfo {author}
  {\bibfnamefont {J.}~\bibnamefont {Mosser}}, \bibinfo {author} {\bibfnamefont
  {C.}~\bibnamefont {Schwartz}}, \bibinfo {author} {\bibfnamefont
  {L.}~\bibnamefont {Garnier}}, \bibinfo {author} {\bibfnamefont
  {F.}~\bibnamefont {Corellou}}, \ and\ \bibinfo {author} {\bibfnamefont
  {F.-Y.}\ \bibnamefont {Bouget}},\ }\href@noop {} {\bibfield  {journal}
  {\bibinfo  {journal} {Plant Physiol.}\ }\textbf {\bibinfo {volume} {144}},\
  \bibinfo {pages} {1360} (\bibinfo {year} {2007})}\BibitemShut {NoStop}%
\bibitem [{\citenamefont {Monnier}\ \emph {et~al.}(2010)\citenamefont
  {Monnier}, \citenamefont {Liverani}, \citenamefont {Bouvet}, \citenamefont
  {Jesson}, \citenamefont {Smith}, \citenamefont {Mosser}, \citenamefont
  {Corellou},\ and\ \citenamefont {Bouget}}]{bouget10}%
  \BibitemOpen
  \bibfield  {author} {\bibinfo {author} {\bibfnamefont {A.}~\bibnamefont
  {Monnier}}, \bibinfo {author} {\bibfnamefont {S.}~\bibnamefont {Liverani}},
  \bibinfo {author} {\bibfnamefont {R.}~\bibnamefont {Bouvet}}, \bibinfo
  {author} {\bibfnamefont {B.}~\bibnamefont {Jesson}}, \bibinfo {author}
  {\bibfnamefont {J.}~\bibnamefont {Smith}}, \bibinfo {author} {\bibfnamefont
  {J.}~\bibnamefont {Mosser}}, \bibinfo {author} {\bibfnamefont
  {F.}~\bibnamefont {Corellou}}, \ and\ \bibinfo {author} {\bibfnamefont
  {F.}~\bibnamefont {Bouget}},\ }\href@noop {} {\bibfield  {journal} {\bibinfo
  {journal} {BMC Genomics}\ }\textbf {\bibinfo {volume} {11}},\ \bibinfo
  {pages} {192} (\bibinfo {year} {2010})}\BibitemShut {NoStop}%
\bibitem [{\citenamefont {Gonze}, \citenamefont {Halloy},\ and\ \citenamefont
  {Goldbeter}(2002)}]{DidierGonze01222002}%
  \BibitemOpen
  \bibfield  {author} {\bibinfo {author} {\bibfnamefont {D.}~\bibnamefont
  {Gonze}}, \bibinfo {author} {\bibfnamefont {J.}~\bibnamefont {Halloy}}, \
  and\ \bibinfo {author} {\bibfnamefont {A.}~\bibnamefont {Goldbeter}},\
  }\href@noop {} {\bibfield  {journal} {\bibinfo  {journal} {Proc. Nat. Acad.
  Sci. USA}\ }\textbf {\bibinfo {volume} {99}},\ \bibinfo {pages} {673}
  (\bibinfo {year} {2002})}\BibitemShut {NoStop}%
\bibitem [{\citenamefont {Barkai}\ and\ \citenamefont
  {Leibler}(2000)}]{barkai00:_circadiannoise}%
  \BibitemOpen
  \bibfield  {author} {\bibinfo {author} {\bibfnamefont {N.}~\bibnamefont
  {Barkai}}\ and\ \bibinfo {author} {\bibfnamefont {S.}~\bibnamefont
  {Leibler}},\ }\href@noop {} {\bibfield  {journal} {\bibinfo  {journal}
  {Nature}\ }\textbf {\bibinfo {volume} {403}},\ \bibinfo {pages} {267}
  (\bibinfo {year} {2000})}\BibitemShut {NoStop}%
\bibitem [{\citenamefont {Pittendrigh}(1954)}]{Pittendrigh54:_comp_temp}%
  \BibitemOpen
  \bibfield  {author} {\bibinfo {author} {\bibfnamefont {C.~S.}\ \bibnamefont
  {Pittendrigh}},\ }\href@noop {} {\bibfield  {journal} {\bibinfo  {journal}
  {Proc. Natl. Acad. Sci. USA}\ }\textbf {\bibinfo {volume} {40}},\ \bibinfo
  {pages} {1018} (\bibinfo {year} {1954})}\BibitemShut {NoStop}%
\bibitem [{\citenamefont {Rensing}\ and\ \citenamefont
  {Ruoff}(2002)}]{Rensing02:_comp_temp}%
  \BibitemOpen
  \bibfield  {author} {\bibinfo {author} {\bibfnamefont {L.}~\bibnamefont
  {Rensing}}\ and\ \bibinfo {author} {\bibfnamefont {P.}~\bibnamefont
  {Ruoff}},\ }\href@noop {} {\bibfield  {journal} {\bibinfo  {journal}
  {Chronobiol. Int.}\ }\textbf {\bibinfo {volume} {19}},\ \bibinfo {pages}
  {807} (\bibinfo {year} {2002})}\BibitemShut {NoStop}%
\bibitem [{\citenamefont {Comas}\ \emph {et~al.}(2008)\citenamefont {Comas},
  \citenamefont {Beersma}, \citenamefont {Hut},\ and\ \citenamefont
  {Daan}}]{M.Comas10012008}%
  \BibitemOpen
  \bibfield  {author} {\bibinfo {author} {\bibfnamefont {M.}~\bibnamefont
  {Comas}}, \bibinfo {author} {\bibfnamefont {D.}~\bibnamefont {Beersma}},
  \bibinfo {author} {\bibfnamefont {R.}~\bibnamefont {Hut}}, \ and\ \bibinfo
  {author} {\bibfnamefont {S.}~\bibnamefont {Daan}},\ }\href@noop {} {\bibfield
   {journal} {\bibinfo  {journal} {J. Biol. Rhythms}\ }\textbf {\bibinfo
  {volume} {23}},\ \bibinfo {pages} {425} (\bibinfo {year} {2008})}\BibitemShut
  {NoStop}%
\bibitem [{\citenamefont {Beersma}, \citenamefont {Daan},\ and\ \citenamefont
  {Hut}(1999)}]{Beersma08011999}%
  \BibitemOpen
  \bibfield  {author} {\bibinfo {author} {\bibfnamefont {D.~G.~M.}\
  \bibnamefont {Beersma}}, \bibinfo {author} {\bibfnamefont {S.}~\bibnamefont
  {Daan}}, \ and\ \bibinfo {author} {\bibfnamefont {R.~A.}\ \bibnamefont
  {Hut}},\ }\href@noop {} {\bibfield  {journal} {\bibinfo  {journal} {J. Biol.
  Rhythms}\ }\textbf {\bibinfo {volume} {14}},\ \bibinfo {pages} {320}
  (\bibinfo {year} {1999})}\BibitemShut {NoStop}%
\bibitem [{\citenamefont {Troein}\ \emph {et~al.}(2009)\citenamefont {Troein},
  \citenamefont {Locke}, \citenamefont {Turner},\ and\ \citenamefont
  {Millar}}]{Troein20091961}%
  \BibitemOpen
  \bibfield  {author} {\bibinfo {author} {\bibfnamefont {C.}~\bibnamefont
  {Troein}}, \bibinfo {author} {\bibfnamefont {J.~C.~W.}\ \bibnamefont
  {Locke}}, \bibinfo {author} {\bibfnamefont {M.~S.}\ \bibnamefont {Turner}}, \
  and\ \bibinfo {author} {\bibfnamefont {A.~J.}\ \bibnamefont {Millar}},\
  }\href@noop {} {\bibfield  {journal} {\bibinfo  {journal} {Current Biology}\
  }\textbf {\bibinfo {volume} {19}},\ \bibinfo {pages} {1961} (\bibinfo {year}
  {2009})}\BibitemShut {NoStop}%
\bibitem [{\citenamefont {Dunlap}(1999)}]{dunlap99:_molec}%
  \BibitemOpen
  \bibfield  {author} {\bibinfo {author} {\bibfnamefont {J.~C.}\ \bibnamefont
  {Dunlap}},\ }\href@noop {} {\bibfield  {journal} {\bibinfo  {journal} {Cell}\
  }\textbf {\bibinfo {volume} {96}},\ \bibinfo {pages} {271} (\bibinfo {year}
  {1999})}\BibitemShut {NoStop}%
\bibitem [{\citenamefont {Young}\ and\ \citenamefont
  {Kay}(2001)}]{young01:_time}%
  \BibitemOpen
  \bibfield  {author} {\bibinfo {author} {\bibfnamefont {M.~W.}\ \bibnamefont
  {Young}}\ and\ \bibinfo {author} {\bibfnamefont {S.}~\bibnamefont {Kay}},\
  }\href@noop {} {\bibfield  {journal} {\bibinfo  {journal} {Nature Genetics}\
  }\textbf {\bibinfo {volume} {2}},\ \bibinfo {pages} {702} (\bibinfo {year}
  {2001})}\BibitemShut {NoStop}%
\bibitem [{\citenamefont {Panda}, \citenamefont {Hogenesch},\ and\
  \citenamefont {Kay}(2002)}]{panda02:_circad}%
  \BibitemOpen
  \bibfield  {author} {\bibinfo {author} {\bibfnamefont {S.}~\bibnamefont
  {Panda}}, \bibinfo {author} {\bibfnamefont {J.~B.}\ \bibnamefont
  {Hogenesch}}, \ and\ \bibinfo {author} {\bibfnamefont {S.~A.}\ \bibnamefont
  {Kay}},\ }\href@noop {} {\bibfield  {journal} {\bibinfo  {journal} {Nature}\
  }\textbf {\bibinfo {volume} {417}},\ \bibinfo {pages} {329} (\bibinfo {year}
  {2002})}\BibitemShut {NoStop}%
\bibitem [{\citenamefont {Forger}\ and\ \citenamefont
  {Peskin}(2003)}]{Forger12092003}%
  \BibitemOpen
  \bibfield  {author} {\bibinfo {author} {\bibfnamefont {D.~B.}\ \bibnamefont
  {Forger}}\ and\ \bibinfo {author} {\bibfnamefont {C.~S.}\ \bibnamefont
  {Peskin}},\ }\href@noop {} {\bibfield  {journal} {\bibinfo  {journal} {Proc.
  Nat. Acad. Sci. USA}\ }\textbf {\bibinfo {volume} {100}},\ \bibinfo {pages}
  {14806} (\bibinfo {year} {2003})}\BibitemShut {NoStop}%
\bibitem [{\citenamefont {Forger}\ \emph {et~al.}(2003)\citenamefont {Forger},
  \citenamefont {{Dean II}}, \citenamefont {Gurdziel}, \citenamefont {Leloup},
  \citenamefont {Lee}, \citenamefont {{Von Gall}}, \citenamefont {Etchegaray},
  \citenamefont {Kronauer}, \citenamefont {Goldbeter}, \citenamefont {Peskin},
  \citenamefont {Jewett},\ and\ \citenamefont {Weaver}}]{forger:_mammalian}%
  \BibitemOpen
  \bibfield  {author} {\bibinfo {author} {\bibfnamefont {D.~B.}\ \bibnamefont
  {Forger}}, \bibinfo {author} {\bibfnamefont {D.~A.}\ \bibnamefont {{Dean
  II}}}, \bibinfo {author} {\bibfnamefont {K.}~\bibnamefont {Gurdziel}},
  \bibinfo {author} {\bibfnamefont {J.-C.}\ \bibnamefont {Leloup}}, \bibinfo
  {author} {\bibfnamefont {C.}~\bibnamefont {Lee}}, \bibinfo {author}
  {\bibfnamefont {C.}~\bibnamefont {{Von Gall}}}, \bibinfo {author}
  {\bibfnamefont {J.-P.}\ \bibnamefont {Etchegaray}}, \bibinfo {author}
  {\bibfnamefont {R.~E.}\ \bibnamefont {Kronauer}}, \bibinfo {author}
  {\bibfnamefont {A.}~\bibnamefont {Goldbeter}}, \bibinfo {author}
  {\bibfnamefont {C.~S.}\ \bibnamefont {Peskin}}, \bibinfo {author}
  {\bibfnamefont {M.~E.}\ \bibnamefont {Jewett}}, \ and\ \bibinfo {author}
  {\bibfnamefont {D.~R.}\ \bibnamefont {Weaver}},\ }\href@noop {} {\bibfield
  {journal} {\bibinfo  {journal} {OMICS}\ }\textbf {\bibinfo {volume} {7}},\
  \bibinfo {pages} {387} (\bibinfo {year} {2003})}\BibitemShut {NoStop}%
\bibitem [{\citenamefont {Locke}\ \emph {et~al.}(2005)\citenamefont {Locke},
  \citenamefont {Southern}, \citenamefont {Kozma-Bognár}, \citenamefont
  {Hibberd}, \citenamefont {Brown}, \citenamefont {Turner},\ and\ \citenamefont
  {Millar}}]{locke05:_extension}%
  \BibitemOpen
  \bibfield  {author} {\bibinfo {author} {\bibfnamefont {J.~C.~W.}\
  \bibnamefont {Locke}}, \bibinfo {author} {\bibfnamefont {M.~M.}\ \bibnamefont
  {Southern}}, \bibinfo {author} {\bibfnamefont {L.}~\bibnamefont
  {Kozma-Bognár}}, \bibinfo {author} {\bibfnamefont {V.}~\bibnamefont
  {Hibberd}}, \bibinfo {author} {\bibfnamefont {P.~E.}\ \bibnamefont {Brown}},
  \bibinfo {author} {\bibfnamefont {M.~S.}\ \bibnamefont {Turner}}, \ and\
  \bibinfo {author} {\bibfnamefont {A.~J.}\ \bibnamefont {Millar}},\
  }\href@noop {} {\bibfield  {journal} {\bibinfo  {journal} {Mol. Syst. Biol.}\
  }\textbf {\bibinfo {volume} {1}},\ \bibinfo {pages} {2005.0013} (\bibinfo
  {year} {2005})}\BibitemShut {NoStop}%
\bibitem [{\citenamefont {Locke}\ \emph {et~al.}(2006)\citenamefont {Locke},
  \citenamefont {Kozma-Bognar}, \citenamefont {Gould}, \citenamefont {Feher},
  \citenamefont {Kevei}, \citenamefont {Nagy}, \citenamefont {Turner},
  \citenamefont {Hall},\ and\ \citenamefont {Millar}}]{locke06:_exper}%
  \BibitemOpen
  \bibfield  {author} {\bibinfo {author} {\bibfnamefont {J.~C.~W.}\
  \bibnamefont {Locke}}, \bibinfo {author} {\bibfnamefont {L.}~\bibnamefont
  {Kozma-Bognar}}, \bibinfo {author} {\bibfnamefont {P.~D.}\ \bibnamefont
  {Gould}}, \bibinfo {author} {\bibfnamefont {B.}~\bibnamefont {Feher}},
  \bibinfo {author} {\bibfnamefont {E.}~\bibnamefont {Kevei}}, \bibinfo
  {author} {\bibfnamefont {F.}~\bibnamefont {Nagy}}, \bibinfo {author}
  {\bibfnamefont {M.~S.}\ \bibnamefont {Turner}}, \bibinfo {author}
  {\bibfnamefont {A.}~\bibnamefont {Hall}}, \ and\ \bibinfo {author}
  {\bibfnamefont {A.~J.}\ \bibnamefont {Millar}},\ }\href@noop {} {\bibfield
  {journal} {\bibinfo  {journal} {Mol. Syst. Biol.}\ }\textbf {\bibinfo
  {volume} {2}},\ \bibinfo {pages} {59} (\bibinfo {year} {2006})}\BibitemShut
  {NoStop}%
\bibitem [{\citenamefont {Zeilinger}\ \emph {et~al.}(2006)\citenamefont
  {Zeilinger}, \citenamefont {Farre}, \citenamefont {Taylor},\ and\
  \citenamefont {Kay}}]{zeilinger06:_arabid_prr7_prr9}%
  \BibitemOpen
  \bibfield  {author} {\bibinfo {author} {\bibfnamefont {M.~N.}\ \bibnamefont
  {Zeilinger}}, \bibinfo {author} {\bibfnamefont {E.~M.}\ \bibnamefont
  {Farre}}, \bibinfo {author} {\bibfnamefont {S.~R.}\ \bibnamefont {Taylor}}, \
  and\ \bibinfo {author} {\bibfnamefont {S.~A.}\ \bibnamefont {Kay}},\
  }\href@noop {} {\bibfield  {journal} {\bibinfo  {journal} {Mol. Syst. Biol.}\
  }\textbf {\bibinfo {volume} {2}},\ \bibinfo {pages} {58} (\bibinfo {year}
  {2006})}\BibitemShut {NoStop}%
\bibitem [{\citenamefont {Salazar}\ \emph {et~al.}(2009)\citenamefont
  {Salazar}, \citenamefont {Saithong}, \citenamefont {Brown}, \citenamefont
  {Foreman}, \citenamefont {Locke}, \citenamefont {Halliday}, \citenamefont
  {Carré}, \citenamefont {Rand},\ and\ \citenamefont
  {Millar}}]{salazar09:_predic}%
  \BibitemOpen
  \bibfield  {author} {\bibinfo {author} {\bibfnamefont {J.~D.}\ \bibnamefont
  {Salazar}}, \bibinfo {author} {\bibfnamefont {T.}~\bibnamefont {Saithong}},
  \bibinfo {author} {\bibfnamefont {P.~E.}\ \bibnamefont {Brown}}, \bibinfo
  {author} {\bibfnamefont {J.}~\bibnamefont {Foreman}}, \bibinfo {author}
  {\bibfnamefont {J.~C.~W.}\ \bibnamefont {Locke}}, \bibinfo {author}
  {\bibfnamefont {K.~J.}\ \bibnamefont {Halliday}}, \bibinfo {author}
  {\bibfnamefont {I.~A.}\ \bibnamefont {Carré}}, \bibinfo {author}
  {\bibfnamefont {D.~A.}\ \bibnamefont {Rand}}, \ and\ \bibinfo {author}
  {\bibfnamefont {A.~J.}\ \bibnamefont {Millar}},\ }\href@noop {} {\bibfield
  {journal} {\bibinfo  {journal} {Cell}\ }\textbf {\bibinfo {volume} {139}},\
  \bibinfo {pages} {1170} (\bibinfo {year} {2009})}\BibitemShut {NoStop}%
\bibitem [{\citenamefont {Fran\c{c}ois}(2005)}]{francois05:_neurospora}%
  \BibitemOpen
  \bibfield  {author} {\bibinfo {author} {\bibfnamefont {P.}~\bibnamefont
  {Fran\c{c}ois}},\ }\href@noop {} {\bibfield  {journal} {\bibinfo  {journal}
  {Biophys. J.}\ }\textbf {\bibinfo {volume} {88}},\ \bibinfo {pages} {2369}
  (\bibinfo {year} {2005})}\BibitemShut {NoStop}%
\bibitem [{\citenamefont {Courties}\ \emph {et~al.}(1994)\citenamefont
  {Courties}, \citenamefont {Vaquer}, \citenamefont {Troussellier},
  \citenamefont {Lautier}, \citenamefont {Chretiennot-Dinet}, \citenamefont
  {Neveux}, \citenamefont {Machado},\ and\ \citenamefont
  {Claustre}}]{Courties94:_ostreo_struc2}%
  \BibitemOpen
  \bibfield  {author} {\bibinfo {author} {\bibfnamefont {C.}~\bibnamefont
  {Courties}}, \bibinfo {author} {\bibfnamefont {A.}~\bibnamefont {Vaquer}},
  \bibinfo {author} {\bibfnamefont {M.}~\bibnamefont {Troussellier}}, \bibinfo
  {author} {\bibfnamefont {J.}~\bibnamefont {Lautier}}, \bibinfo {author}
  {\bibfnamefont {M.~J.}\ \bibnamefont {Chretiennot-Dinet}}, \bibinfo {author}
  {\bibfnamefont {J.}~\bibnamefont {Neveux}}, \bibinfo {author} {\bibfnamefont
  {M.~C.}\ \bibnamefont {Machado}}, \ and\ \bibinfo {author} {\bibfnamefont
  {H.}~\bibnamefont {Claustre}},\ }\href@noop {} {\bibfield  {journal}
  {\bibinfo  {journal} {Nature}\ }\textbf {\bibinfo {volume} {370}},\ \bibinfo
  {pages} {255} (\bibinfo {year} {1994})}\BibitemShut {NoStop}%
\bibitem [{\citenamefont {Chretiennot-Dinet}\ \emph {et~al.}(1995)\citenamefont
  {Chretiennot-Dinet}, \citenamefont {Courties}, \citenamefont {Vaquer},
  \citenamefont {Neveux}, \citenamefont {Claustre}, \citenamefont {Lautier},\
  and\ \citenamefont {Machado}}]{chretiennot95:_ostreo_struc}%
  \BibitemOpen
  \bibfield  {author} {\bibinfo {author} {\bibfnamefont {M.~J.}\ \bibnamefont
  {Chretiennot-Dinet}}, \bibinfo {author} {\bibfnamefont {C.}~\bibnamefont
  {Courties}}, \bibinfo {author} {\bibfnamefont {A.}~\bibnamefont {Vaquer}},
  \bibinfo {author} {\bibfnamefont {J.}~\bibnamefont {Neveux}}, \bibinfo
  {author} {\bibfnamefont {H.}~\bibnamefont {Claustre}}, \bibinfo {author}
  {\bibfnamefont {J.}~\bibnamefont {Lautier}}, \ and\ \bibinfo {author}
  {\bibfnamefont {M.~C.}\ \bibnamefont {Machado}},\ }\href@noop {} {\bibfield
  {journal} {\bibinfo  {journal} {Phycologia}\ }\textbf {\bibinfo {volume}
  {4}},\ \bibinfo {pages} {285­292} (\bibinfo {year} {1995})}\BibitemShut
  {NoStop}%
\bibitem [{\citenamefont {{E. Derelle \emph{et
  al.}}}(2006)}]{derelle06:_genome}%
  \BibitemOpen
  \bibfield  {author} {\bibinfo {author} {\bibnamefont {{E. Derelle \emph{et
  al.}}}},\ }\href@noop {} {\bibfield  {journal} {\bibinfo  {journal} {Proc.
  Nat. Acad. Sci. USA}\ }\textbf {\bibinfo {volume} {103}},\ \bibinfo {pages}
  {11647} (\bibinfo {year} {2006})}\BibitemShut {NoStop}%
\bibitem [{\citenamefont {Djouani-Tahri}\ \emph {et~al.}(2010)\citenamefont
  {Djouani-Tahri}, \citenamefont {Motta}, \citenamefont {Bouget},\ and\
  \citenamefont {Corellou}}]{corellou10}%
  \BibitemOpen
  \bibfield  {author} {\bibinfo {author} {\bibfnamefont {E.}~\bibnamefont
  {Djouani-Tahri}}, \bibinfo {author} {\bibfnamefont {J.-P.}\ \bibnamefont
  {Motta}}, \bibinfo {author} {\bibfnamefont {F.-Y.}\ \bibnamefont {Bouget}}, \
  and\ \bibinfo {author} {\bibfnamefont {F.}~\bibnamefont {Corellou}},\
  }\href@noop {} {\bibfield  {journal} {\bibinfo  {journal} {Plant Signal
  Behav.}\ }\textbf {\bibinfo {volume} {5}},\ \bibinfo {pages} {332} (\bibinfo
  {year} {2010})}\BibitemShut {NoStop}%
\bibitem [{\citenamefont {Alabadi}\ \emph {et~al.}(2001)\citenamefont
  {Alabadi}, \citenamefont {Oyama}, \citenamefont {Yanovsky}, \citenamefont
  {Harmon}, \citenamefont {Mas},\ and\ \citenamefont
  {Kay}}]{DavidAlabadi08032001}%
  \BibitemOpen
  \bibfield  {author} {\bibinfo {author} {\bibfnamefont {D.}~\bibnamefont
  {Alabadi}}, \bibinfo {author} {\bibfnamefont {T.}~\bibnamefont {Oyama}},
  \bibinfo {author} {\bibfnamefont {M.~J.}\ \bibnamefont {Yanovsky}}, \bibinfo
  {author} {\bibfnamefont {F.~G.}\ \bibnamefont {Harmon}}, \bibinfo {author}
  {\bibfnamefont {P.}~\bibnamefont {Mas}}, \ and\ \bibinfo {author}
  {\bibfnamefont {S.~A.}\ \bibnamefont {Kay}},\ }\href@noop {} {\bibfield
  {journal} {\bibinfo  {journal} {Science}\ }\textbf {\bibinfo {volume}
  {293}},\ \bibinfo {pages} {880} (\bibinfo {year} {2001})}\BibitemShut
  {NoStop}%
\bibitem [{\citenamefont {Gates}\ and\ \citenamefont {DeLuca}(1975)}]{Gates75}%
  \BibitemOpen
  \bibfield  {author} {\bibinfo {author} {\bibfnamefont {B.}~\bibnamefont
  {Gates}}\ and\ \bibinfo {author} {\bibfnamefont {M.}~\bibnamefont {DeLuca}},\
  }\href@noop {} {\bibfield  {journal} {\bibinfo  {journal} {Arch Biochem
  Biophys}\ }\textbf {\bibinfo {volume} {169}},\ \bibinfo {pages} {616}
  (\bibinfo {year} {1975})}\BibitemShut {NoStop}%
\bibitem [{\citenamefont {Lefranc}, \citenamefont {Hennequin},\ and\
  \citenamefont {Glorieux}(1992)}]{lefranc92:_correl_dimen}%
  \BibitemOpen
  \bibfield  {author} {\bibinfo {author} {\bibfnamefont {M.}~\bibnamefont
  {Lefranc}}, \bibinfo {author} {\bibfnamefont {D.}~\bibnamefont {Hennequin}},
  \ and\ \bibinfo {author} {\bibfnamefont {P.}~\bibnamefont {Glorieux}},\
  }\href@noop {} {\bibfield  {journal} {\bibinfo  {journal} {Phys. Lett. A}\
  }\textbf {\bibinfo {volume} {163}},\ \bibinfo {pages} {269} (\bibinfo {year}
  {1992})}\BibitemShut {NoStop}%
\bibitem [{\citenamefont {Letellier}\ and\ \citenamefont
  {Aguirre}(2010)}]{letellier10:_inter}%
  \BibitemOpen
  \bibfield  {author} {\bibinfo {author} {\bibfnamefont {C.}~\bibnamefont
  {Letellier}}\ and\ \bibinfo {author} {\bibfnamefont {L.~A.}\ \bibnamefont
  {Aguirre}},\ }\href@noop {} {\bibfield  {journal} {\bibinfo  {journal} {Phys.
  Rev. E}\ }\textbf {\bibinfo {volume} {82}},\ \bibinfo {pages} {016204}
  (\bibinfo {year} {2010})}\BibitemShut {NoStop}%
\bibitem [{\citenamefont {Gutenkunst}\ \emph {et~al.}(2007)\citenamefont
  {Gutenkunst}, \citenamefont {Waterfall}, \citenamefont {Casey}, \citenamefont
  {Brown}, \citenamefont {Myers},\ and\ \citenamefont {Sethna}}]{Gutenkunst07}%
  \BibitemOpen
  \bibfield  {author} {\bibinfo {author} {\bibfnamefont {R.~N.}\ \bibnamefont
  {Gutenkunst}}, \bibinfo {author} {\bibfnamefont {J.~J.}\ \bibnamefont
  {Waterfall}}, \bibinfo {author} {\bibfnamefont {F.~P.}\ \bibnamefont
  {Casey}}, \bibinfo {author} {\bibfnamefont {K.~S.}\ \bibnamefont {Brown}},
  \bibinfo {author} {\bibfnamefont {C.~R.}\ \bibnamefont {Myers}}, \ and\
  \bibinfo {author} {\bibfnamefont {J.~P.}\ \bibnamefont {Sethna}},\
  }\href@noop {} {\bibfield  {journal} {\bibinfo  {journal} {PLoS Comput.
  Biol.}\ }\textbf {\bibinfo {volume} {3}},\ \bibinfo {pages} {e189} (\bibinfo
  {year} {2007})}\BibitemShut {NoStop}%
\bibitem [{\citenamefont {Kurosawa}\ and\ \citenamefont
  {Isawa}(2002)}]{Kurosawa02:_satur}%
  \BibitemOpen
  \bibfield  {author} {\bibinfo {author} {\bibfnamefont {G.}~\bibnamefont
  {Kurosawa}}\ and\ \bibinfo {author} {\bibfnamefont {Y.}~\bibnamefont
  {Isawa}},\ }\href@noop {} {\bibfield  {journal} {\bibinfo  {journal} {J. Biol
  Rhythms}\ }\textbf {\bibinfo {volume} {17}},\ \bibinfo {pages} {568}
  (\bibinfo {year} {2002})}\BibitemShut {NoStop}%
\bibitem [{\citenamefont {Wong}, \citenamefont {Tsai},\ and\ \citenamefont
  {Liao}(2007)}]{Wong07:_prot_deg}%
  \BibitemOpen
  \bibfield  {author} {\bibinfo {author} {\bibfnamefont {W.~W.}\ \bibnamefont
  {Wong}}, \bibinfo {author} {\bibfnamefont {T.~Y.}\ \bibnamefont {Tsai}}, \
  and\ \bibinfo {author} {\bibfnamefont {J.~C.}\ \bibnamefont {Liao}},\
  }\href@noop {} {\bibfield  {journal} {\bibinfo  {journal} {{Mol. Syst.
  Biol.}}\ }\textbf {\bibinfo {volume} {{3}}},\ \bibinfo {pages} {{130}}
  (\bibinfo {year} {{2007}})}\BibitemShut {NoStop}%
\bibitem [{\citenamefont {Morant}\ \emph {et~al.}(2009)\citenamefont {Morant},
  \citenamefont {Thommen}, \citenamefont {Lemaire}, \citenamefont
  {Vandermoere}, \citenamefont {Parent},\ and\ \citenamefont
  {Lefranc}}]{degradator}%
  \BibitemOpen
  \bibfield  {author} {\bibinfo {author} {\bibfnamefont {P.-E.}\ \bibnamefont
  {Morant}}, \bibinfo {author} {\bibfnamefont {Q.}~\bibnamefont {Thommen}},
  \bibinfo {author} {\bibfnamefont {F.}~\bibnamefont {Lemaire}}, \bibinfo
  {author} {\bibfnamefont {C.}~\bibnamefont {Vandermoere}}, \bibinfo {author}
  {\bibfnamefont {B.}~\bibnamefont {Parent}}, \ and\ \bibinfo {author}
  {\bibfnamefont {M.}~\bibnamefont {Lefranc}},\ }\href@noop {} {\bibfield
  {journal} {\bibinfo  {journal} {Phys. Rev. Lett.}\ }\textbf {\bibinfo
  {volume} {102}},\ \bibinfo {pages} {068104} (\bibinfo {year}
  {2009})}\BibitemShut {NoStop}%
\bibitem [{\citenamefont {Mather}\ \emph {et~al.}(2009)\citenamefont {Mather},
  \citenamefont {Bennett}, \citenamefont {Hasty},\ and\ \citenamefont
  {Tsimring}}]{Mather:_tsimringPRL}%
  \BibitemOpen
  \bibfield  {author} {\bibinfo {author} {\bibfnamefont {W.}~\bibnamefont
  {Mather}}, \bibinfo {author} {\bibfnamefont {M.~R.}\ \bibnamefont {Bennett}},
  \bibinfo {author} {\bibfnamefont {J.}~\bibnamefont {Hasty}}, \ and\ \bibinfo
  {author} {\bibfnamefont {L.~S.}\ \bibnamefont {Tsimring}},\ }\href@noop {}
  {\bibfield  {journal} {\bibinfo  {journal} {{Phys. Rev. Lett.}}\ }\textbf
  {\bibinfo {volume} {{102}}},\ \bibinfo {pages} {{068105}} (\bibinfo {year}
  {{2009}})}\BibitemShut {NoStop}%
\end{thebibliography}
\end{document}